%% file: main.tex
\documentclass[sigconf, nonacm]{acmart}

\usepackage{microtype}
\usepackage{graphicx}
\usepackage{booktabs} 
\newcommand{\ceil}[1]{\left \lceil #1 \right \rceil }
\usepackage{amsmath}
\usepackage{multirow}
\usepackage{algorithm}
\usepackage[noend]{algpseudocode}
\usepackage{enumitem}
\usepackage{caption}
\usepackage{subcaption}

\usepackage{array}

\newcolumntype{C}[1]{>{\centering\let\newline\\\arraybackslash\hspace{0pt}}m{#1}}

\begin{document}
\pagestyle{plain}
\title{Incremental IVF Index Maintenance for Streaming Vector Search}

\author{Jason Mohoney}
\affiliation{
\institution{University of Wisconsin-Madison}
}

\author{Anil Pacaci}
\affiliation{
\institution{Apple}
}

\author{Shihabur Rahman Chowdhury}
\affiliation{
\institution{Apple}
}

\author{Umar Farooq Minhas}
\affiliation{
\institution{Apple}
}

\author{Jeffrey Pound}
\affiliation{
\institution{Apple}
}

\author{Cedric Renggli}
\affiliation{
\institution{Apple}
}

\author{Nima Reyhani}
\affiliation{
\institution{Apple}
}

\author{Ihab F. Ilyas}
\affiliation{
\institution{Apple}
}

\author{Theodoros Rekatsinas}
\affiliation{
\institution{Apple}
}

\author{Shivaram Venkataraman}
\affiliation{
\institution{University of Wisconsin-Madison}
}
\renewcommand{\shortauthors}{Mohoney, et al.}
\begin{abstract}

\input{s0_abstract}
\end{abstract}

\maketitle

\input{s1_intro}

\input{s2_background}

\input{s3_ada_ivf}

\input{s4_quality_maintenance}

\input{s6_experiments}

\input{s7_discussion}

\input{s8_conclusion}


\bibliographystyle{ACM-Reference-Format}
\bibliography{ref}

\end{document}

%% file: s0_abstract.tex
The prevalence of vector similarity search in modern machine learning applications and the continuously changing nature of data processed by these applications necessitate efficient and effective index maintenance techniques for vector search indexes.
Designed primarily for static workloads, existing vector search indexes degrade in search quality and performance as the underlying data is updated unless costly index reconstruction is performed.
To address this, we introduce Ada-IVF, an incremental indexing methodology for Inverted File (IVF) indexes. Ada-IVF consists of 1) an adaptive maintenance policy that decides which index partitions are problematic for performance and should be repartitioned and 2) a local re-clustering mechanism that determines how to repartition them. 
Compared with state-of-the-art dynamic IVF index maintenance strategies, Ada-IVF achieves an average of $2\times$ and up to $5\times$ higher update throughput across a range of benchmark workloads.

%% file: s1_intro.tex
\section{Introduction}

Modern machine learning applications increasingly rely on high-dimensional vector embeddings to transform complex data such as images, text, or entities in knowledge graphs to vector representations that retain semantically meaningful information~\cite{gordo2016deep, cai2018comprehensive, hossain2019comprehensive, minaee2021deep, JMLR:v23:20-852}. 
Vector similarity search is a critical component in such applications as it enables more accurate and contextualized search~\cite{grbovic2018real, haldar2019applying, hashemi2021neural, qin2021mixer} and recommendations~\cite{okura2017embedding, liu2017related, wang2018billion, pal2020pinnersage, liu2022monolith} over multi-modal data.
Partitioned indexes for vector similarity search have recently gained widespread adoption in these modern machine learning applications due to their performance and scalability \cite{zhang_fast_nodate, jiang_co-design_2023, mohoney_high-throughput_2023, guo_accelerating_2020, xu_spfresh_2023, wei2020analyticdb}. A common type of partitioned index is the Inverted File (IVF) index, where a vector quantization algorithm (typically k-means) 
is used to partition a vector dataset, and the resulting clusters constitute the partitions of the index~\cite{jegou_product_2011}.
While existing IVF index implementations are designed for static workloads, i.e., the partitioning is performed based on the initial state of the underlying vector dataset, real-world deployments require high-throughput \emph{search} and \emph{updates} over \emph{dynamic vector data} where the underlying vector dataset is continuously modified through insertions and deletions \cite{xu_spfresh_2023, singh2021freshdiskann, baranchuk_dedrift_2023}. In addition to being robust against modifications to the vector dataset, the indexes must be robust against changing query patterns over time, as observed in ~\cite{baranchuk_dedrift_2023}. 
In this work, we study the effect of updates on the IVF index's search and update throughput and propose an incremental maintenance methodology for IVF indexes.

IVF indexes out-of-the-box do not have the notion of inserting new vectors or deleting existing vectors once constructed. Indeed, the most common method used by practitioners today is to rebuild the index from scratch to reflect any updates that have accumulated over time. However, depending on the scale of the vector dataset and the volume and frequency of updates, a full index rebuild can be prohibitively expensive. For example, it takes multiple days to rebuild an IVF index from scratch for billion-scale vector datasets~\cite{xu_spfresh_2023, chen_spann_nodate}, making it necessary to revisit how updates can be reflected. Devising such an update mechanism consists of re-adjusting the partitioning of the high-dimensional space defined by the clusters and ensuring that the re-adjusted partitioning: (i) keeps the reconstruction error, \emph{i.e.}, the average distance between a vector and its nearest cluster centroid, at minimum, otherwise, queries would require scanning more partitions to reach a target recall and degrades search throughput; (ii) does not create an imbalance in the size distribution of the partitions, which can result in variable search latency across queries~\cite{baranchuk_dedrift_2023, xu_spfresh_2023, jegou_improving_2010}.

In industrial vector search workloads, we observe that the partition access patterns of search and update operations over an IVF index are non-uniform and change over time. 
For instance, in a typical day of a KG entity search workload we studied~\cite{mohoney_high-throughput_2023, ilyas2022saga}, we find that only $15\%$ of partitions were accessed during search operations, and $80\%$ of the updates affected partitions that were not accessed by any search operation.
Such skewed access patterns induced by real-world vector search workloads present an opportunity for efficient and effective maintenance of IVF indexes over dynamic datasets. 
Specifically, index maintenance overhead can be minimized by devising a local, incremental indexing strategy and focusing the maintenance process on frequently accessed partitions during search. To the best of our knowledge, no existing approaches in the literature utilize partition access patterns for IVF index maintenance.  

To this end, we propose Ada-IVF, an incremental maintenance mechanism for IVF indexes. Our approach is based on (i) the observation that reconstruction error and partition imbalance serve as indicators of an IVF index's search quality and performance (Section \ref{sec:ivf}), and (ii) workload patterns can be utilized for local, incremental maintenance of its underlying partitions (Section \ref{sec:observations}). Ada-IVF consists of a workload-adaptive \emph{policy} that identifies which partitions should be reindexed based on real-time statistics in order to minimize reconstruction error and partition imbalance and a \emph{mechanism} that performs local reindexing over the target subset of partitions. Our policy uses \emph{global} and \emph{partition-local} indicator functions that quantify changes in reconstruction error and partition imbalance. In addition, Ada-IVF tracks read frequencies for individual partitions to quantify a \emph{temperature} of each partition, which is then used by the local indicator function. Using temperature allows Ada-IVF to prioritize the maintenance of commonly accessed partitions and to avoid unnecessary work on rarely accessed partitions. Through its incremental maintenance policy and local reindexing mechanism, Ada-IVF effectively mitigates IVF index performance degradation due to updates.

Our experimental analysis confirms the efficacy of our approach. We conduct a comparison of Ada-IVF against the state-of-the-art LIRE \cite{xu_spfresh_2023} and other baseline IVF maintenance methodologies on synthetic data and public benchmarks. We show that Ada-IVF obtains similar or better search throughput when compared to baseline methods with an average of $2\times$ improvement in throughput across all workloads and a maximum of $5\times$ update throughput improvement on synthetic data. Furthermore, we verify Ada-IVF's robustness across various workload patterns using traces of industrial workloads and public and synthetic benchmarks, thereby establishing its applicability in diverse real-world scenarios. 

In summary, our contributions include:
\begin{itemize}[noitemsep,topsep=0pt]
\item Global and local indicator functions to quantify the impact of updates on global index and individual partition levels.
\item A workload-adaptive policy that uses real-time statistics and these indicator functions to monitor reconstruction error and partition imbalance and decide when to perform reindexing.
\item A local reindexing mechanism that uses k-means to split and reindex candidates and their neighboring partitions.
\item A comprehensive empirical evaluation demonstrating the impact of updates on IVF indexes under a diverse array of real and synthetic workloads.
\end{itemize}

In Section \ref{sec:background}, we discuss the preliminaries behind our study, detailing existing index solutions and systems as well as observations on real workloads. Section \ref{sec:design} provides an overview of the design of Ada-IVF. Section \ref{sec:iqm} details our contributions to IVF index quality maintenance. Section \ref{sec:experiments} presents our experimental results on internal industrial workloads and public and synthetic benchmarks, with a suite of microbenchmarks to validate our contributions. Section \ref{sec:discussion} concludes with a discussion of our findings and related work.

%% file: s2_background.tex
\section{Preliminaries \& Motivation}
\label{sec:background}

We now define the streaming vector search problem. We also provide a brief overview of IVF indexes, their use of vector search, and existing methods for supporting updates. Finally, we discuss our observations on the characteristics of streaming vector search workloads and their impacts on the IVF index performance, which motivate the incremental indexing strategy introduced in this paper.

\subsection{Vector Search \& Data Updates}

Streaming vector search \cite{aguerrebere_locally-adaptive_2024, singh2021freshdiskann} is the process of finding the top-k nearest neighbors of a $d$-dimensional query vector $q$ in an evolving set of $d$-dimensional vectors $\mathbf{X}$. We consider the following operations over the set of vectors $\mathbf{X}$:
\begin{itemize}
    \item \textbf{Insertion}: Adding a new vector $x$ to the set \(\mathbf{X}\), making \(\mathbf{X} = \mathbf{X} \cup \{x\}\).
    \item \textbf{Deletion}: Removing an existing vector \(x \in \mathbf{X}\) from the set, resulting in \( = \mathbf{X} \setminus \{x\}\).
    \item \textbf{Search}: Find the top-k nearest neighbors to $q$ in $\mathbf{X}$ according to a similarity metric.
\end{itemize}

Obtaining the exact top-k nearest neighbors of a query vector $q$ in $\mathbf{X}$ is prohibitively expensive if $|\mathbf{X}|$ is large. In practice, approximate nearest neighbor (ANN) search methods that leverage indexes are used, providing a trade-off between accuracy and performance. Search accuracy is most commonly measured using \emph{recall}, defined as \(r = |\mathbf{G} \cap \mathbf{R}| / k\), where $\mathbf{R}$ is the set of ids returned from approximate search and $\mathbf{G}$ is the set of ground truth ids obtained by a linear scan. Performance, on the other hand, is most generally measured as the number of queries processed per second (QPS). Indexes for ANN search need to be updated otherwise performance can deteriorate with respect to a fixed recall target.

A streaming vector search workload is an ordered set of search and update operations over the set of vectors $\mathbf{X}$. As defined above, search operations correspond to finding k-nearest neighbors of a query vector over the current state of $\mathbf{X}$, and update operations modify set $\mathbf{X}$. These operations can be batched, searching, or updating multiple vectors at a time. 

\subsection{IVF Indexes}
\label{sec:ivf}

An IVF index is a partitioned data structure for high-dimensional vector search that consists of $n_c$ partitions (clusters) and their representative vectors (centroids). An IVF index over a vector dataset $\mathbf{X}$ is constructed offline using a vector quantizer, typically a variant of k-means \cite{jegou_improving_2010, chen_spann_nodate}; consequently, the index construction process requires a global view of the underlying vector dataset $\mathbf{X}$.
Clustering-based partitioning of the dataset enables similar vectors (based on the embedding distance) to be assigned to the same partition.
At search time, such partitioning is utilized to reduce the number of distance computations performed; 
the distance from the query vector $q$ to each cluster centroid is computed to identify $n_p$ nearest partitions, and each of the $p$ partitions is scanned to obtain the approximate k-nearest neighbors. By controlling the subset partitions scanned for each query, $n_p$ provides a way to navigate the trade-off between quality and performance.

\begin{figure*}[!htb]
\minipage{0.32\textwidth}
    \centering
    \includegraphics[width=\textwidth]{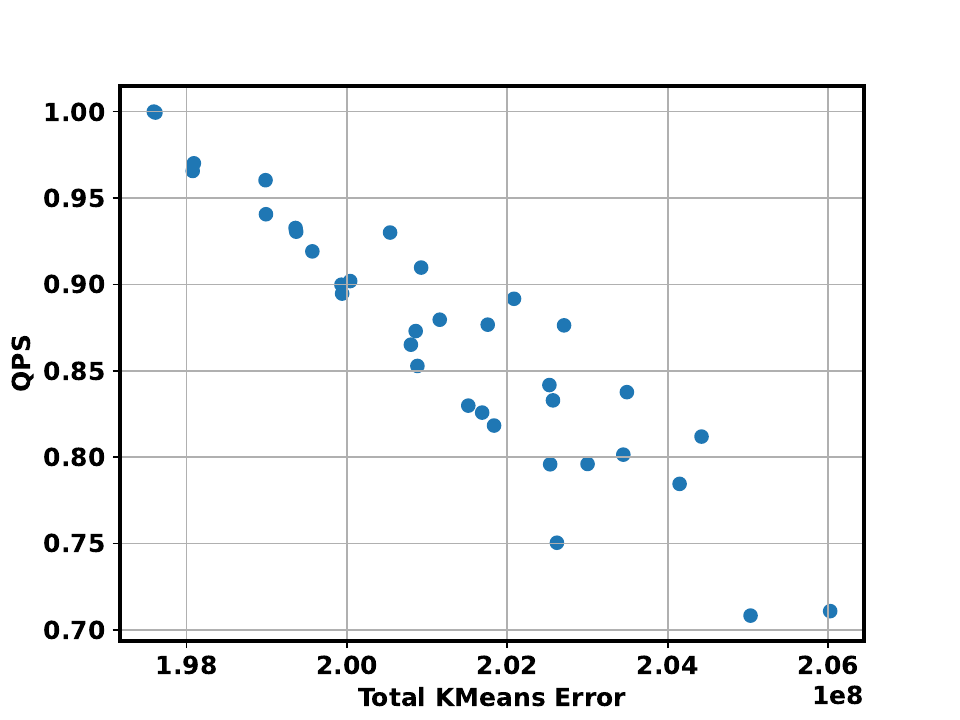}
    \caption{Static IVF indexes trained with balanced k-means on SIFT1M. Each point is a different initialization or number of iterations for k-means. As error increases, QPS degrades for recall@0.9}
    \label{fig:overlap}
\endminipage\hfill
\minipage{0.32\textwidth}
    \centering
    \includegraphics[width=\textwidth]{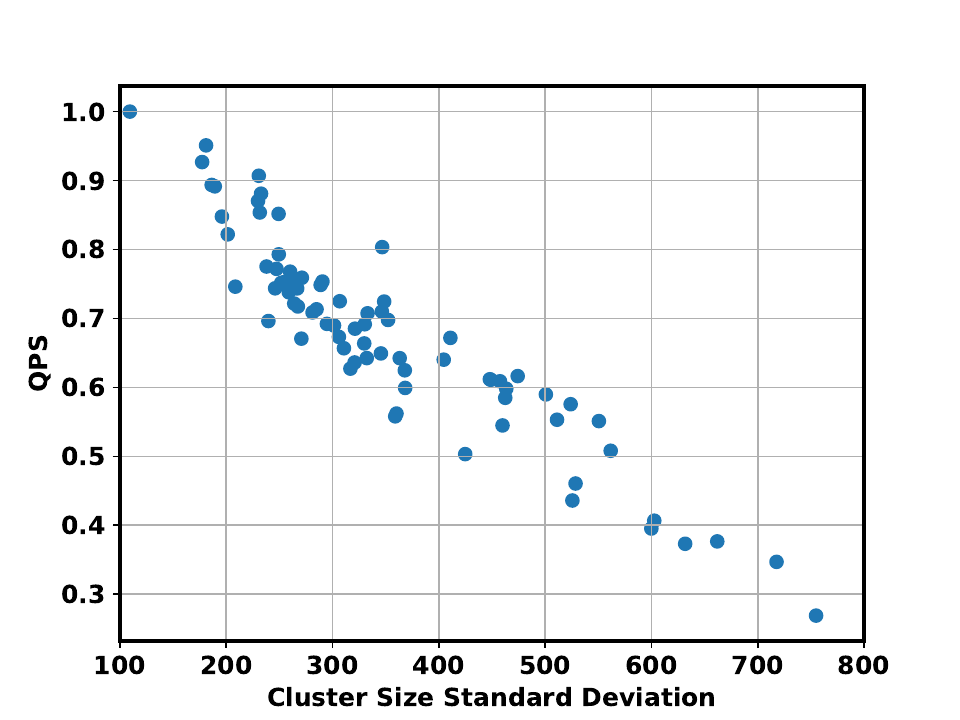}
    \caption{Evaluation of a \emph{Frozen} IVF index on a SIFT1M dynamic workload with insert/delete ratio = 1. As partition imbalance increases, read throughput degrades for recall@0.9}
    \label{fig:imbalance}
\endminipage\hfill
\minipage{0.32\textwidth}%
    \centering
    \includegraphics[width=.95\textwidth]{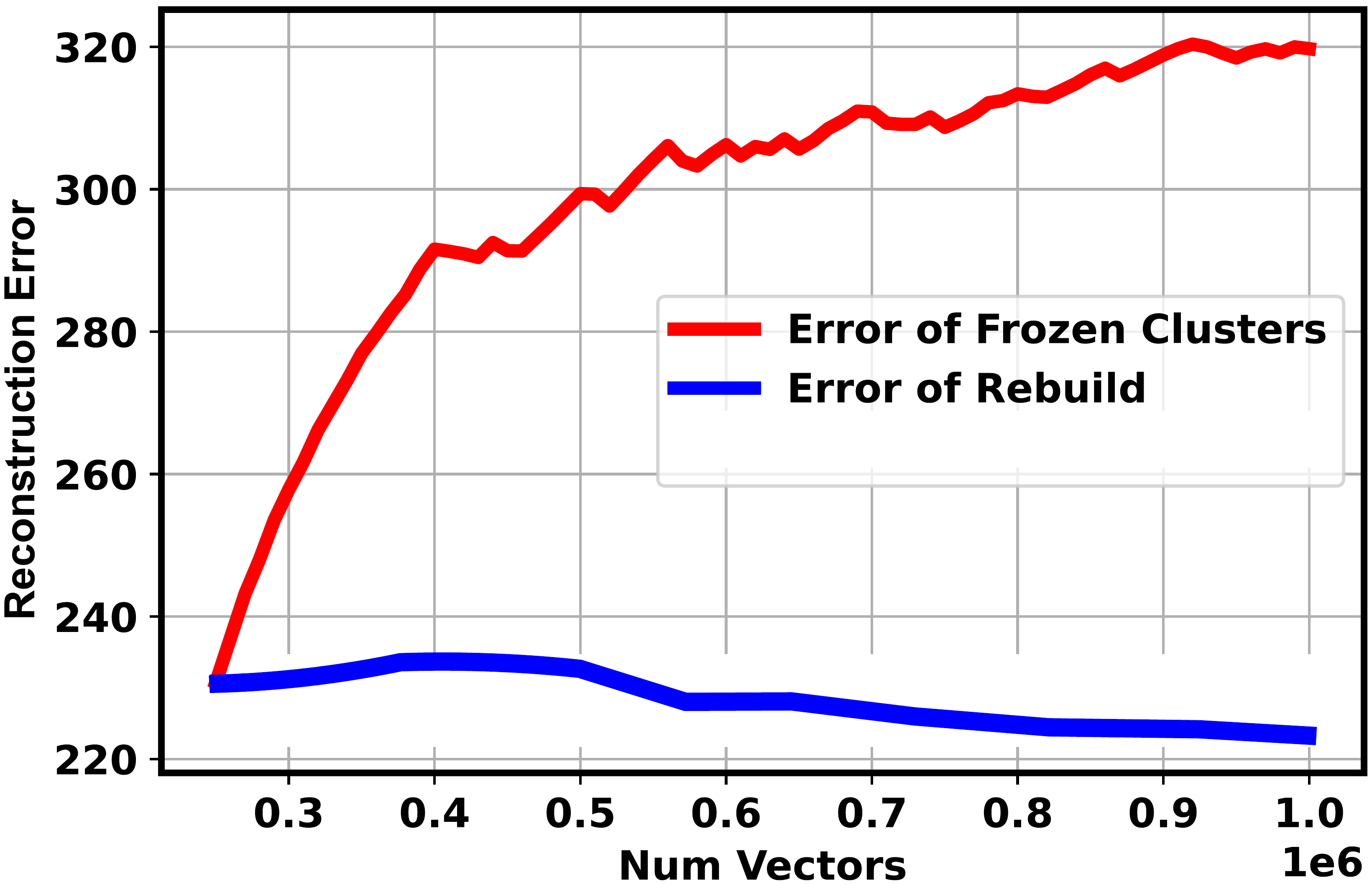}
    \caption{Evaluation of an IVF index reconstruction error as vectors from SIFT1M are inserted into the index. The error grows away from the error observed from fully rebuilding the index.}
    \label{fig:growing_error}
\endminipage
\end{figure*}



The primary indicators of IVF index search performance are i) \emph{partition imbalance} and ii) \emph{reconstruction error}. Partition imbalance refers to the uneven size distribution of partitions, and it adversely affects the search throughput as queries will take longer to scan overpopulated partitions. A common strategy to mitigate partition imbalance during index construction is to use balanced k-means, which can produce roughly equal-sized partitions so that latency is consistent across queries for a given $n_p$ \cite{xu_spfresh_2023}. The reconstruction error is the average distance from each vector to its nearest centroid, and the objective function of the quantization algorithm minimizes the error. Numerous works have been proposed to minimize the reconstruction error of static vector search indexes \cite{ge_optimized_2014, wu_multiscale_2017, babenko_tree_2015}. Indexes with a larger reconstruction error have less compact clusters and require queries to scan more partitions to find their nearest neighbors, degrading search performance. Figure \ref{fig:overlap} measures the search throughput vs. reconstruction error (k-means error) on the SIFT1M dataset for IVF indexes trained using various configurations of balanced k-means with a fixed number of centroids ($n_c$=1000). We observe degradation in performance as the error increases. Thus, partition imbalance and reconstruction error must be minimized to optimize search performance in IVF indexes.

Naively modifying partitions of the IVF index via partition append and deletes can support update operations in IVF indexes. However, as the clustering of vectors deviates from its initial state, such modifications can lead to an increase in partition imbalance and reconstruction error, which results in degradation in search quality and performance (See Section \ref{sec:observations}). To ensure robust search performance in the face of updates, it is necessary to address the increase in partition imbalance and reconstruction error. Below are previously proposed maintenance strategies for supporting updates in IVF indexes.

\begin{itemize}
    \item \textbf{Rebuild} \cite{wei2020analyticdb}: Periodically rebuilds the index from scratch.
    \item \textbf{Frozen} \cite{noauthor_pgvectorpgvector_nodate}: Partitions are modified without changing the centroids or number of partitions.
    \item \textbf{Update Centroids} \cite{arandjelovic2013all}: Similar to \emph{Frozen}, but centroids are updated upon partition modification to reflect the true mean of the partition.
    \item \textbf{DeDrift} \cite{baranchuk_dedrift_2023}: Similar to \emph{Update Centroids}, but for each update, the $k_1 << n_c$ largest and smallest partitions are collected and reclustered using k-means, keeping the total number of partitions fixed.
    \item \textbf{LIRE} \cite{xu_spfresh_2023}: Uses splitting/merging of partitions that violate pre-defined size thresholds. The contents of violating partitions with $r_c$ neighboring partitions are reassigned to minimize reconstruction error. Neighboring partitions are determined by centroid distance to the violating partition.
\end{itemize}


Each IVF index maintenance approach presents a trade-off between the maintenance overhead (efficiency) and the ability to address partition imbalance and reconstruction error (effectiveness). While \emph{Rebuild} is highly effective, i.e., it can minimize both imbalance and error by building the IVF index from scratch over the latest state of the dataset. Yet, it is inefficient and has the highest maintenance overhead. Although it can be acceptable for applications with static or slowly changing vector datasets, \emph{Rebuild} is prohibitively expensive for large-scale vector datasets used in real-world applications, taking days for billion-scale datasets \cite{xu_spfresh_2023}.

\emph{Frozen}, at the other end of the efficiency vs effectiveness trade-off, does not incur any maintenance overhead. However, it is only effective when the distribution of vectors and queries is uniform. As described in the next section in detail, \emph{Frozen} does not address the increasing partition imbalance or reconstruction error for real-world workloads with skewed, non-uniform vector and query distributions. \emph{Update Centroids} incurs the cost of modifying centroids with each update, which mitigates increasing reconstruction error but does not address imbalance. \emph{DeDrift} not only updates centroids but also pays the cost of reclustering a subset of the partitions for each update operation, effectively addressing both imbalance and reconstruction error. \emph{LIRE}, the state-of-the-art incremental IVF indexing solution, focuses on addressing imbalance, and its overhead mainly stems from the reassignment phase, which varies with the number of violating partitions. Overall, these existing update methods for IVF indexes do not consider the trade-off between maintenance overhead and search performance (i.e., partition imbalance and reconstruction error) in a holistic manner.

\begin{figure}[]
    \centering
    \includegraphics[width=.5\textwidth]{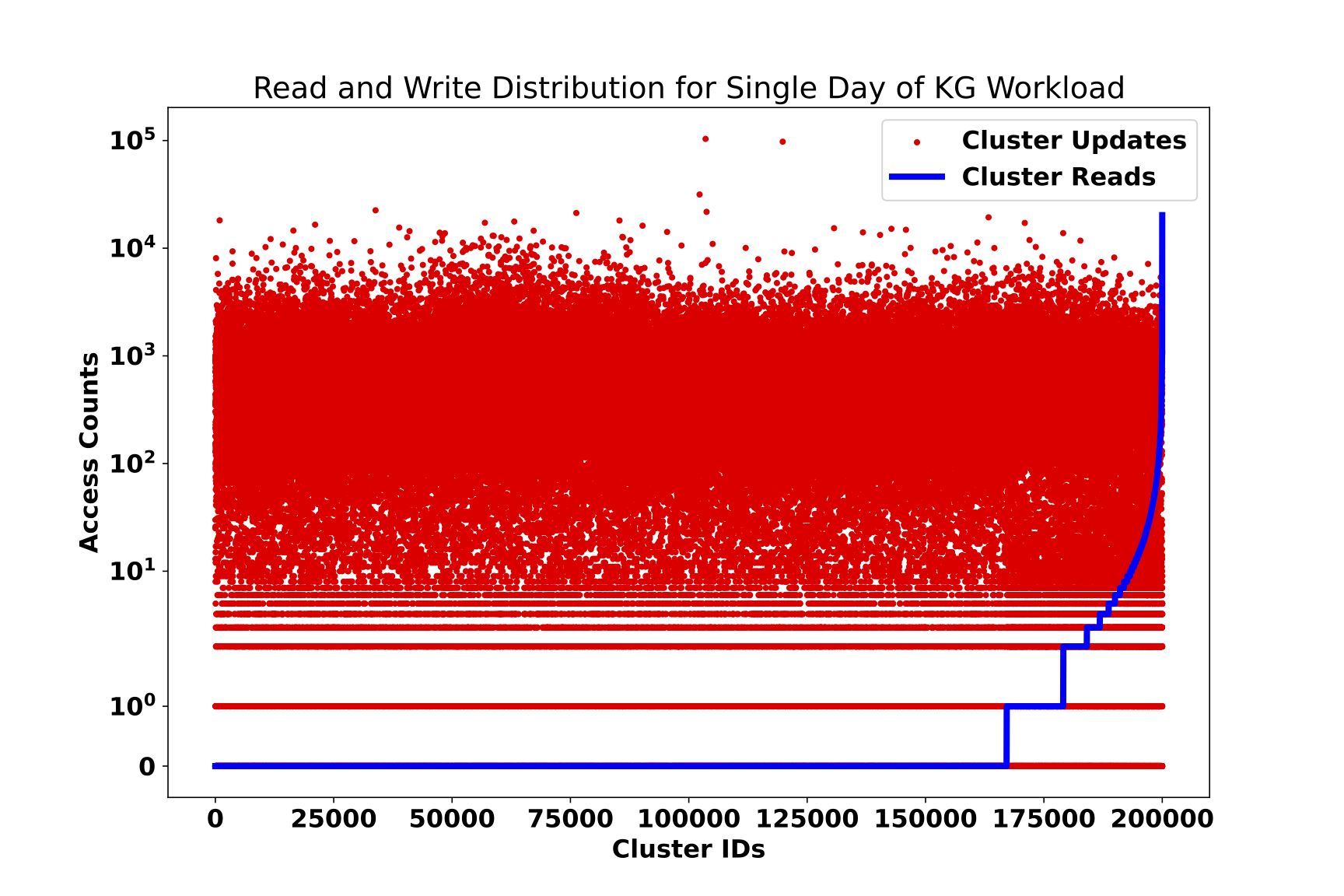}
    \caption{Read and write access patterns of partitions in an internal entity search workload. Partition IDs are ordered by their read count. The read distribution is skewed and uncorrelated with the write distribution, showing that many partitions are modified but not read from.}
    \label{fig:access_patterns}
\end{figure}

\begin{figure}[]
    \centering
    \includegraphics[width=.5\textwidth]{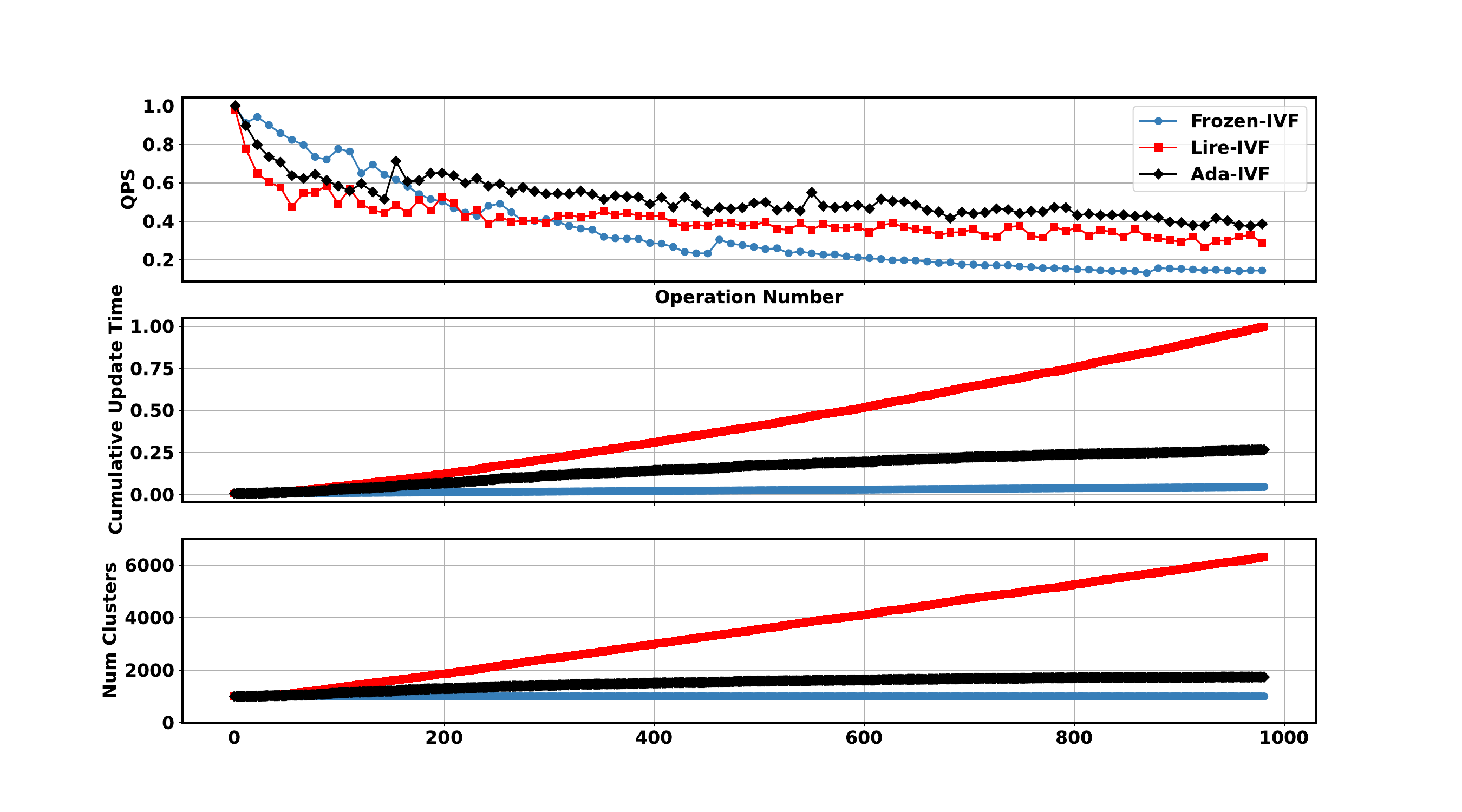}
    \caption{Evaluation of LIRE and Ada-IVF for a dynamic MSTuring10m workload where queries are localized to a few partitions. LIRE and our locality-aware approach Ada-IVF achieve similar QPS, but LIRE creates $3\times$ as many partitions and requires $4\times$ the update time due to its maintenance of partitions that are not accessed by the queries. }
    \label{fig:overtrigger}
\end{figure}

\subsection{Impact of Updates on IVF Indexes}
\label{sec:observations}

Here, we motivate Ada-IVF and its incremental maintenance methodology by discussing the characteristics of vector search workloads and the impact of updates on existing IVF index maintenance strategies. First, we show how partition imbalance and reconstruction error are impacted by simple baseline strategies such as \emph{Frozen}, which does not change the clustering. Second, we measure partition access patterns of an industrial search workload to show the skew in read and write access patterns of real-world workloads. Finally, we show how the state-of-the-art incremental IVF indexing approach, \emph{LIRE}, incurs unnecessary maintenance overhead by over-triggering reindexing for rarely accessed partitions.


Figure \ref{fig:imbalance} demonstrates the relationship between the search performance and partition imbalance by showing the query throughput and the standard deviation of partition sizes over different snapshots of an IVF index updated via a sequence of insertions and deletions from the SIFT1M dataset. As vectors are added to and deleted from the index, partitions grow imbalanced, as measured by the increasing standard deviation of partition sizes. As the imbalance grows, there is a clear correlation to degrading search performance as more vectors are scanned to achieve the recall target. We observe in Figure \ref{fig:growing_error} that the reconstruction error increases with updates on a similar insert-only workload. The error for \emph{Frozen} deviates from the error observed for \emph{Rebuild} as the number of vectors added to the index increases. Increasing reconstruction error results in degrading search performance (Figure \ref{fig:overlap}). Our measurements demonstrate that IVF indexes exhibit increasing imbalance and reconstruction error due to updates, necessitating a maintenance strategy.

Next, we examine an industrial search workload over knowledge graph embeddings \cite{ilyas2022saga} and find that partitions are accessed non-uniformly by search and update operations, exhibiting skewed partition access locality. 
In Figure \ref{fig:access_patterns}, we plot the total number of reads and writes for each partition over an IVF index for a single day of the workload.
We see that $15\%$ of partitions are accessed for search operations in a given day, with only $3\%$ accessed by more than ten queries. We also see for this workload that the search and update operations access different sets of partitions, with $80\%$ of updates affecting partitions that have not been accessed by search operations and $96\%$ affecting partitions that are accessed by fewer than ten search queries. As we describe in the following sections in detail, these locality patterns provide an optimization opportunity for efficient and effective incremental update of IVF indexes compared to existing IVF update methodologies.

Finally, we observe that existing IVF index maintenance methods do not account for partition read locality. This results in DeDrift and LIRE over-triggering index maintenance for partitions that are rarely accessed. In Figure \ref{fig:overtrigger}, we compare the update performance of LIRE against our locality-aware approach using a workload where queries are localized to a small number of partitions and updates modify all partitions uniformly. While both approaches can achieve similar search QPS, LIRE spends $4\times$ more time processing updates. This is because LIRE eagerly reindexes partitions that exceed the pre-defined size threshold regardless of whether recent queries access them, while our proposed approach lazily reindexes partitions as queries access them. 

%% file: s3_ada_ivf.tex
\begin{figure*}[ht!]
    \centering
    \includegraphics[width=0.9\textwidth]{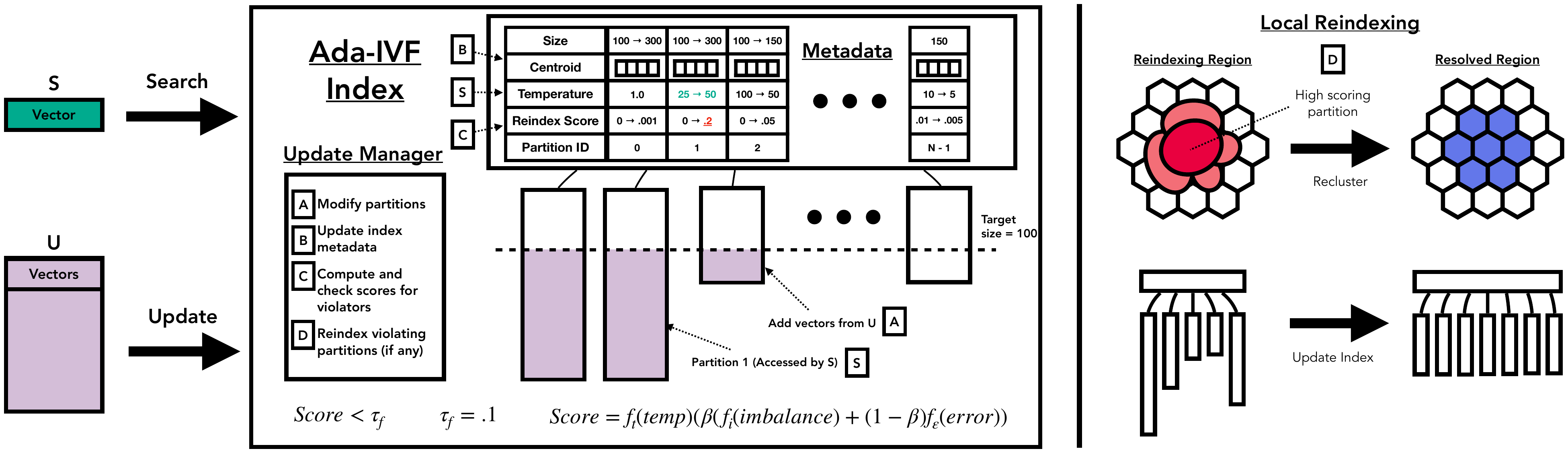}
    \caption{Ada-IVF Overview and Example.
    Search and update operations modify the partitions (A) and metadata (S, B). For example, S increases the temperature of partition 1 from 25 to 50, and U increases the size from 100 to 300. After each update, a reindexing score is computed for modified partitions (C), and if the score violates a threshold, the partition is selected for reindexing (D).}
    \label{fig:architecture}
\end{figure*}

\section{Ada-IVF Overview}
\label{sec:design}

We introduce Ada-IVF, an incremental maintenance solution for IVF indexes. Figure \ref{fig:architecture} depicts the high-level overview of our incremental reindexing solution. In brief, Ada-IVF maintains robust search performance in the face of updates by selectively reindexing a subset of partitions that negatively contribute to the partition imbalance and reconstruction error, proxies for index performance, as discussed in the previous section. Ada-IVF has the following steps:

\begin{itemize}
    \item Step 1: Maintain index metadata. Ada-IVF maintains the size, centroid, and \emph{read temperature} of partitions. Where the temperature denotes the frequency and recency of reads for a given partition. Search operations modify the temperature (\textbf{S}), and update operations modify the partition contents, size, and centroid (\textbf{U, A, B}).
    \item Step 2: Monitor index partitions for increases in imbalance and reconstruction error. Partitions are assigned a \emph{reindexing score} (\textbf{C}), which is a function of imbalance and reconstruction error, a higher score indicating higher imbalance/error. Partitions exceeding a score threshold of $\tau_f$ are \emph{violators} (e.g., Partition 1).
    \item Step 3: Reindex violating partitions using local reindexing. Violating partitions are split and merged with neighboring partitions (\textbf{D}) to minimize imbalance and reconstruction error.
\end{itemize}

\paragraph{Tracking Index Metadata}
Ada-IVF tracks each partition's size, initial and current centroid vector, read temperature, and reindexing score as index metadata. The partition size determines imbalance by checking the deviation from a target partition size. The change in the centroid from its initial state signals an increase in reconstruction error. The read \emph{temperature} is a float that captures how often a partition is accessed by a search operation. The \emph{reindexing score} is a float indicating whether a partition should be reindexed. The total memory overhead is the index metadata stored by Ada-IVF is $(2 \times n_c \times d + 3) \times 4$ bytes for an index with $n_c$ partitions, and $d$ is the vector dimension.

It is important that the index metadata is maintained in a lightweight process to avoid slowdowns in search and update operations. Consider the effect of a search operation S followed by an update U on the Ada-IVF index in Figure \ref{fig:architecture}. The search operation S modifies the temperature for each partition, which is a small overhead compared to executing the search operation. The temperature is increased for partitions accessed by S and decreased for all non-accessed partitions. For example, partition $1$ is accessed by S, and its temperature increases from 25 to 50 while the other partitions decrease correspondingly. The update operation U modifies the size, centroid, and reindexing score of a set of partitions ($0$, $1$, $2$ in the example). Updating the size is made inexpensive by maintaining a counter for each partition. The centroids of modified partitions are updated incrementally rather than computing the mean over all vectors in the partition. Then, the reindexing score is computed using the new size and centroid of the modified partitions. We observed that the metadata tracking process can take up to $20\%$ of the maintenance time of Ada-IVF, where the majority of the time is due to updating centroids. The metadata update process is detailed in \ref{sec:tracking}. 

\paragraph{Update Manager} Ada-IVF's update manager (Algorithm \ref{alg:reindex}) consists of a reindexing policy that identifies which and when partitions should be reindexed and a mechanism for reindexing them.

The reindexing policy is informed by two indicator functions: \emph{local} and \emph{global} indicator functions, where both functions are evaluated after each update. The \emph{local} indicator function $f(c)$ computes a reindexing score indicating if a specific partition $c$ is imbalanced or has drifted from its initial state and is described in Section \ref{sec:local_reindexing}. If the indicator function exceeds a given threshold, the update manager identifies the partition as a violator and a candidate for reindexing. For example, partition $1$ has a score of .2, which exceeds the threshold of .1, and therefore the partition is selected for reindexing. The \emph{global} indicator function $G(I)$ acts as a fail-safe procedure to ensure that our local reindexing actions don't degrade imbalance or reconstruction error globally. The entire index is rebuilt if the \emph{global} indicator exceeds a threshold. The global indicator function is described in \ref{sec:global_reindex}. Indicator function values are checked after a batch of updates are issued to the index.

The local reindexing mechanism resolves quality violations by splitting violating partitions and merging with neighboring partitions using a variant of the original clustering algorithm. The violating partition is split using balanced k-means to meet the target partition size, e.g.,  partition $1$ in Figure \ref{fig:architecture} is split into three partitions of size 100. Then, the merging phase balances the selected partitions and minimizes the error introduced by the splitting phase. Here, the nearest $r_c$ partitions to the violator are selected based on centroid distance. The split partitions and neighboring partitions are then iteratively refined using a clustering algorithm.
$r_c$ enables Ada-IVF to control the trade-off between reindexing cost and resulting index quality, a larger value of $r_c$ selects a larger region and incurs a more expensive reindexing cost but improves the resulting index quality. The reindexing mechanism and the impact of $r_c$ is further detailed in Section \ref{sec:local_mechanism}.

\begin{algorithm} 
\caption{Reindexing Manager} \label{alg:reindex}
\begin{flushleft} 
\textbf{Input} IVF Index $I$, target partition size $\tau_s$, local indicator threshold $\tau_f$, reclustering radius $r_c$, global indicator threshold $\tau_G$  \\
\textbf{Output} Updated Index $I'$ \\
\end{flushleft} 
\begin{algorithmic}[1]
\Function{CheckReindex}{$I$}
    \State $I'$ = $I$
    \State $\mathbf{C'} = \{\}$
    \For{$c \in $\ I.Partitions} // check for local violations
        \If {$f(c) > \tau_f$}
            \State $\mathbf{C'} = \mathbf{C'} \cup \{c\}$
        \EndIf
    \EndFor
    \If {$|\mathbf{C'}| > 0$} 
        \State $I'$ = LocalReindex($\mathbf{I'}, \mathbf{C'}$, $\tau_s$, $r_c$)
    \EndIf
    \If {$G(I') > \tau_G$} // check for global violation
        \State $I'$ = BuildIVF($I'$)
    \EndIf
    \State \Return $I'$
\EndFunction
\end{algorithmic}
\end{algorithm}

%% file: s4_quality_maintenance.tex
\section{Index Quality Maintenance} \label{sec:iqm}

This section describes the details of Ada-IVF's incremental index maintenance. Beginning with the update rules for index metadata, covering local reindexing, and concluding with global reindexing.

\subsection{Tracking Index Properties}
\label{sec:tracking}
To identify the subset of partitions that cause degradation in query throughput, it is necessary to track how the underlying clustering properties evolve over time.  For this purpose, we maintain the set of vectors that are present in each partition ($\mathbf{X_i}$) and the corresponding mean of the vectors ($\mathbf{\mu_i}$) for each partition. These are used to determine if the reconstruction error has increased due to modifications of the partition. We also track the partition read temperature ($T_i)$, which captures the frequency and recency of reads (scans) performed over the vectors in the partition.

At the global index level, we track the standard deviation of partition sizes ($\sigma$) to help us detect when partitions are imbalanced and the global reconstruction error ($\varepsilon$) to determine if the clustering quality has degraded globally. We do not use temperature at the global level. 
Table \ref{tab:cluster_properties} summarizes the set of properties tracked by Ada-IVF. 

\begin{table}[h]
\centering
\begin{tabular}{l p{6.5cm}}
\hline
\textbf{Symbol} & \textbf{Description} \\
\hline
\(\sigma\) & Standard deviation of partition size \\
\(\varepsilon\) & Reconstruction error (MSE) of vectors and assigned centroid \\
\cmidrule{1-2}
\(\mathbf{X_i}\) & Set of vectors in partition $i$ \\

\(T_i\) & Partition read temperature  \\
\( \boldsymbol{\mu}_i \) & running centroid of the partition \\
\hline
\end{tabular}
\caption{Tracked Clustering Properties}
\label{tab:cluster_properties}
\end{table}

Updating each partition's local properties is shown in Algorithm \ref{alg:update_cluster_properties} and Algorithm \ref{alg:update_temp}.
Algorithm \ref{alg:update_cluster_properties} shows how vectors in a partition (Lines 2-5) and its running centroid (Lines 6-8) are updated when a batch of vectors $X_{\delta}$ are added or removed from the partition.

Read temperature $T_i$ is a floating point value defined between $[1.0, \inf)$ that indicates the frequency and recency of reads for a given partition. The intuition behind tracking the temperature for a partition is that, for workloads with skewed access patterns, as we observed from real-world applications (Section \ref{sec:observations}), degradations of partitions that are frequently accessed have a larger impact on the overall workload performance. Therefore, Ada-IVF prioritizes partitions with high temperature (frequently accessed partitions) during reindexing. Temperature is updated for all partitions for each search operation issued to the system using the Algorithm \ref{alg:update_temp}. If a given query accesses a partition, its temperature increases (Lines 5-7) according to a multiplicative heating factor $\eta$; if not, the temperature is decreased by cooling factor $\nu$ (Line 9). 

\begin{algorithm} 
\caption{Partition Update Rule} \label{alg:update_cluster_properties}
\begin{flushleft} 
\textbf{Input} partition vectors $\mathbf{X_i}$, partition size, mean $\mu_i$, delta vectors $X_\delta$, $isDelete$ \\
\end{flushleft}
\begin{flushleft} 
\textbf{Output} updated partition and properties 
$\mathbf{X_{i+1}}$, $\mu_{i+1}$
\end{flushleft}
\begin{algorithmic}[1] 
\Function{UpdatePartitionProperties}{$\mathbf{X_i}, \mu_i, \mathbf{X_\delta}, isDelete$}
    \If {\(isDelete\)}
        \State \(s_\delta  = -|\mathbf{X_\delta}|\); \(\mathbf{X_{i+1}} = \mathbf{X_i} - \mathbf{X_\delta}\)
    \Else
        \State\(s_{\delta} = |\mathbf{X_\delta}|\); \(\mathbf{X_{i+1}} = \mathbf{X_i} \cup \mathbf{X_\delta}\)
    \EndIf
    \State \(\mu_\delta = Mean(\mathbf{X_\delta})\); \(s_i = |\mathbf{X_i}|\)
    \State \( s_{i+1} = s_i + s_\delta \)  
    \State \( \mu_{i+1} = \mu_{i} + \dfrac{s_\delta}{ s_{i+1}}(\mu_\delta - \mu_{i}) \)  
    \State \Return \(\mathbf{X_{i+1}}, \mu_{i+1} \)
\EndFunction
\end{algorithmic}
\end{algorithm}

\begin{algorithm}
\caption{Single Query Temperature Update} \label{alg:update_temp}
\begin{flushleft} 
\textbf{Input} Query vector $q$, set of index partitions $\mathbf{C}$, nprobe $n_p$, heating parameter $\eta$, cooling parameter $\nu$ \\
\end{flushleft}
\begin{flushleft} 
\textbf{Output} Updated temperature for each partition $\mathbf{C}.T$
\end{flushleft}
\begin{algorithmic}[1]
\Function{UpdateTemperature}{$q, \mathbf{C}, n_p, \eta, \nu$}
    \State $\mathbf{C_{knn}}$, $\mathbf{C_d}$ = KNN($q$, $\mathbf{C}.\mu$, $k$=$n_p$)
    \State \(\mathbf{C_d} = \dfrac{\mathbf{C_d}[0]}{\mathbf{C_d}} \) // scale by distance to nearest centroid
    \For{each \( c \in \mathbf{C}\)}
        \If{\(c \in \mathbf{C_{knn}}\)}
            \State \(d_\mu = \mathbf{C_d}[c]\)
            \State \( c.T = c.T(1 + d_\mu \eta) \)
        \Else
            \State \( min(c.T = c.T(1 - \nu), 1.0) \)
        \EndIf
    \EndFor
    \State \Return \(\mathbf{C}.T\)
\EndFunction
\end{algorithmic}
\end{algorithm}

\subsection{Local Reindexing} \label{sec:local_reindexing}

Next, we introduce Ada-IVF's local indicator function $f$, which is used to identify partitions that should be considered as candidates for reindexing and describe its local reindexing algorithm for addressing the increase in imbalance and reconstruction error without rebuilding the entire index from scratch.

\subsubsection{Local Indicator Function $f$}
The local indicator function ($f$) quantifies the deviation from a partition's original state w.r.t. imbalance and reconstruction error, determining when a local reindexing operation should be triggered. This function is evaluated for a single partition upon modification of the partition. The indicator function captures changes in the partition's size and mean over time, and the partition temperature is used as a scaling factor.
If the value of $f$ for a given partition exceeds a threshold, the partition is denoted as a \textit{violator} and is selected for reclustering. Deviations in partition size introduce imbalance in the index, while change in the partition mean is a proxy for increasing reconstruction error. 

Partition temperature is used to determine the severity of deviations; degradations in a partition with a high temperature have a significant impact on the overall query throughput for a workload because more queries access that partition. Therefore, we should more aggressively select highly accessed partitions for reindexing. Conversely, a partition with a low temperature (i.e., rarely or never accessed) has a negligible effect on overall index performance. By using temperature in its indicator function, Ada-IVF can perform index maintenance \emph{lazily}, i.e., when needed due to negative impact on index performance, instead of performing index maintenance \emph{eagerly} at all times. As will be shown experimentally in Section \ref{sec:experiments}, this significantly reduces index maintenance time with minimal impact on index performance.  

Taking these factors into account, the local indicator function $f$ for a given partition $c$ is defined as:

\begin{equation} \label{eqn:local_indicator}
\begin{split}
    f(c, c_0) = f_T(c.T) (\beta f_s(c.s, \tau_s) + (1 - \beta) f_d(c, c_0)) \\
\end{split}
\end{equation}

The individual functions $f_T$, $f_s$, and $f_d$ control the contribution of the partition temperature, deviation in size, and drift from the initial partition state $c_0$ to the current state $c$. $\beta \in [0, 1]$ controls the contribution of imbalance vs. drift.
If the value of $f$ exceeds the threshold $\tau_f$, the corresponding partition is chosen for reindexing. We next discuss how we design $f_T$, $f_s$, and $f_d$.

\paragraph{Temperature Scaling Function $f_T$}
The function $f_T$ aims to capture how frequently we should re-index a partition based on the temperature $T$ of the partition. Thus, if the temperature is high, it should have a larger score, and vice-versa for low temperature. Thus, we use a linear function parameterized by a scale factor $\alpha$
\begin{equation}
    f_T(T) = \alpha T
\end{equation}

\paragraph{Local Size Imbalance Function $f_s$}
The function $f_s$ captures if a partition has grown too large or small. Thus, denoting $\tau_s$ as the target partition size, we define $f_s$ as:

\begin{equation}
f_s(s, \tau_s) =
\left\{
    \begin{array}{lr}
        (s - \tau_s) / \tau_s, & \text{if } s \geq \tau_s \\
        (\tau_s - s) / s, & \text{if } s < \tau_s
    \end{array}
\right\}
\end{equation}

\paragraph{Local Drift Function $f_d$}

The function $f_d$ captures if a partition $c$ has drifted from its initial state $c_0$, which is a proxy for increasing reconstruction error. We quantify drift by measuring the relative change in partition mean $\mu$ from initial $\mu_0$.

\begin{equation}
    f_d = ||(\boldsymbol{\mu} - \boldsymbol{\mu_0})|| / ||\boldsymbol{\mu_0}||
\end{equation}

\begin{figure}[]
    \centering
    \includegraphics[width=.45\textwidth]{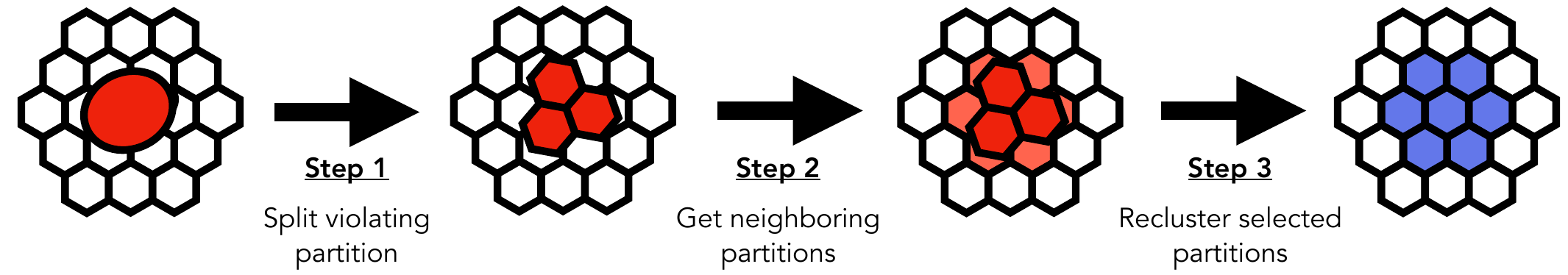}
    \caption{Local reindexing mechanism. Violating partitions are split to meet a target size and merged with neighboring partitions using balanced k-means to minimize imbalance and reconstruction error.}
    \label{fig:local_reindexing_diagram}
\end{figure}

\begin{algorithm} 
\caption{Local Reindexing} \label{alg:local_mechanism}
\textbf{Input} IVF Index $I$, Violating partitions $C'$, target partition size $\tau_s$, reclustering radius $r_c$, and local kmeans iterations $\iota$ \\
\begin{flushleft}
\textbf{Output} updated IVF Index $I'$
\end{flushleft}
\begin{algorithmic}[1]
\Function{LocalReindex}{$I$, $C'$, $\tau_s$, $r_c$, $i$}

\State $I'$ = $I$; $\mathbf{X}$ = \{\}; $\mathbf{M}$ = \{\}

\For{each \( c \in C' \)} 
    \If{$|c| > \tau_s$} // Step 1: split violating partitions
        \State $\mathbf{S} = KMeans(c.X, k=\ceil{|\mathbf{c.X}\| / \tau_s}, \iota)$
        \State $\mathbf{M} = \mathbf{M} \cup \mathbf{S.\mu}$
    \Else 
        \State $\mathbf{M} = \mathbf{M} \cup \{c.\mu\}$
    \EndIf
    \State $\mathbf{X} = \mathbf{X} \cup c.\mathbf{X}$; $I'$ = $I'$.remove($c$)
\EndFor

// Step 2: get neighboring partitions
\State $\mathbf{R} = KNN(\mathbf{M}, I'.\mu, k=r_c)$  

\For{each \(c \in \mathbf{R}\)}
    \State $\mathbf{X} = \mathbf{X} \cup c.X$; $\mathbf{M} = \mathbf{M} \cup \{c.\mu\}$; $I'$ = $I'$.remove($c$)
\EndFor

// Step 3: recluster partitions with initialized centroids 
\State $\mathbf{C_n}$ = KMeans($\mathbf{X}$, initial\_centroids=$\mathbf{M}$, $\iota$) 

\For{each \( c \in \mathbf{C_n} \)}: $I'$.add($c$)
\EndFor

\State return $I'$
\EndFunction
\end{algorithmic}
\end{algorithm}

\subsubsection{Local Reindexing Algorithm}
\label{sec:local_mechanism}

Local reindexing consists of three phases: i) splitting and deleting of violating partitions, ii) finding the nearest partitions to the violating partitions, and iii) applying k-means to the vectors of the violating partitions and their neighbors (Algorithm \ref{alg:local_mechanism}). Partitions are selected as violators if the value of their indicator function $f$ exceeds the threshold $\tau_f$. If the size of the violating partition exceeds the target partition size $\tau_s$, it is split using k-means (Lines 4-6). Otherwise, it is deleted (Line 9). The centroids of split and deleted partitions are used in KNN search to find their corresponding neighboring partitions (Line 10). The reindexing radius $r_c$ controls how many neighboring partitions to consider for each violator, where a larger value will reduce reconstruction error while increasing the reindexing time. In our experiments, we used a fixed value of $r_c=25$ to balance between reindexing time and minimizing error. A potential improvement of our method is to set $r_c$ individually for each violator based on centroid distance, but we have not implemented this approach. Balanced k-means is applied to the region for $\iota$ iterations and is initialized using the previous centroid state $M$ (Line 13). Finally, the newly created partitions are added to the index.

\subsection{Global Reindexing} \label{sec:global_reindex}

While Ada-IVFs local reindexing mechanism can handle deviations in individual partitions when updates and their impact on partitions exhibit locality, it is possible to observe deviations in a large number of partitions or overall index quality that might render local, partition-specific maintenance ineffective. It is necessary for Ada-IVF to identify such an index state and perform a full index rebuild. Here, we describe the global indicator function $G$ that we use to determine when to reconstruct the index from scratch instead of performing local reindexing. Similar to local reindexing, we use a \emph{global} indicator function to quantify deviations in the overall clustering and construct a new index over the current state of $\mathbf{X}$. 

The global indicator function $G$ is a function of the current reconstruction error $\varepsilon$ of the clustering, the partition size standard deviation $\sigma$, and the estimated reconstruction error $\varepsilon'$, the estimation of the reconstruction error of k-means clustering if a full index rebuild were to occur over the current set of vectors. 

The function $G$ is defined as:
\begin{equation} \label{eqn:global_indicator}
\begin{split}
    G(\varepsilon, \sigma, \varepsilon') = \gamma G_s(\sigma) + (1 - \gamma) G_d(\varepsilon, \varepsilon') \leq \tau_G \\
\end{split}
\end{equation}

The individual functions $G_s$, $G_\varepsilon$ capture the contribution of the size imbalance and reconstruction error, and $\gamma$ controls their relative contribution; $\tau_G$ is a tunable threshold.

\paragraph{Global Imbalance Function $G_s$}

We use the relative change in the standard deviation of partition sizes to estimate the increase in global imbalance. Intuitively, the standard deviation in partition size indicates how imbalanced the partitions are with respect to a balanced partitioning. The change in standard deviation captures if the imbalance has increased due to updates performed on the index.  

\begin{equation}
    G_s = |\sigma - \sigma_0| / \sigma_0 
\end{equation}

\begin{figure}[]
    \centering
    \includegraphics[width=.4\textwidth]{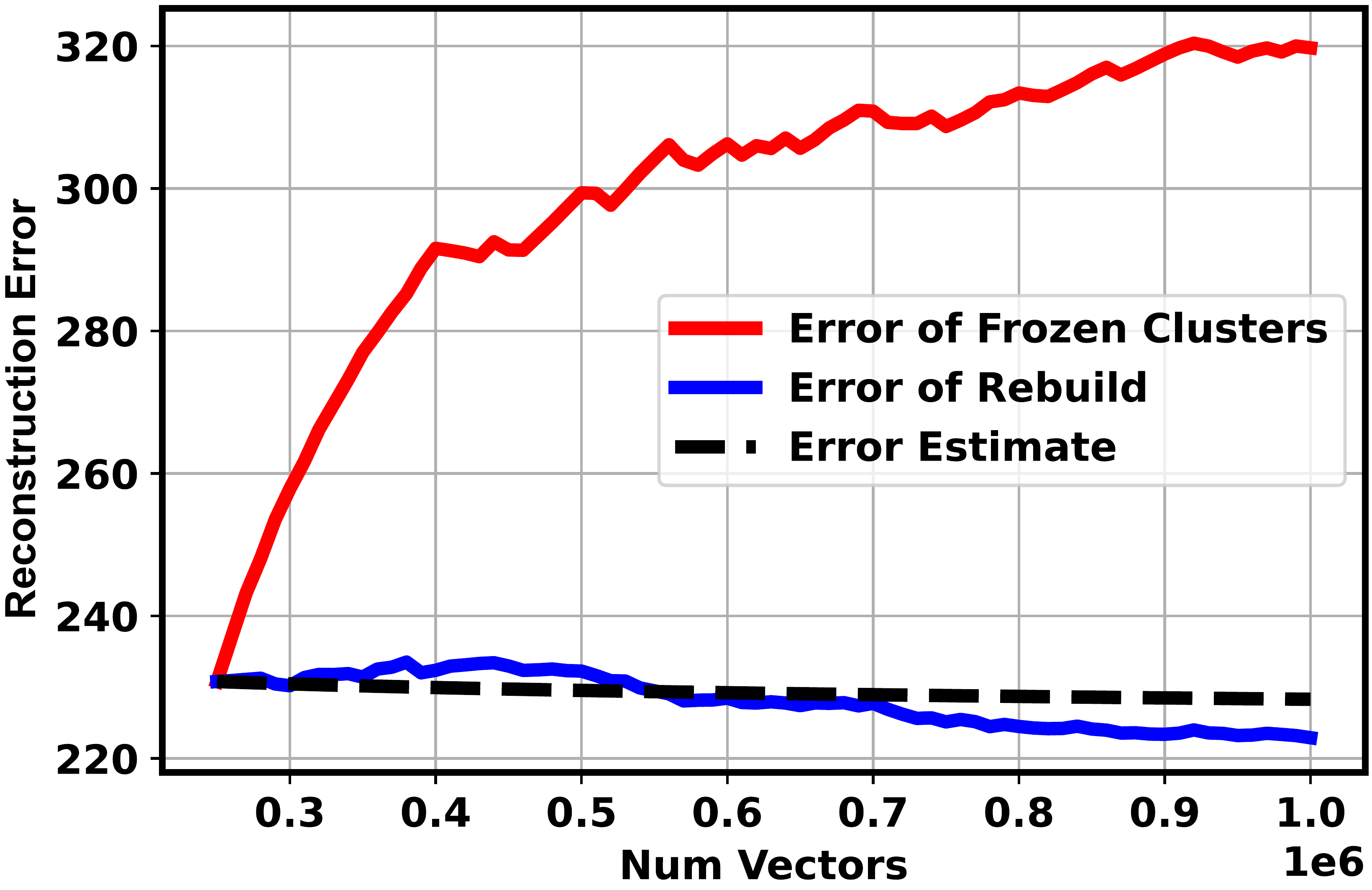}
    \caption{Error estimate $\varepsilon'$ relative to true error for a growing index on Sift1M. Error of frozen clustering grows while the estimate is within $2.5$\% of that obtained by a full rebuild.}
    \label{fig:sift1m_error}
\end{figure}

\begin{figure}[]
    \centering
    \includegraphics[width=.4\textwidth]{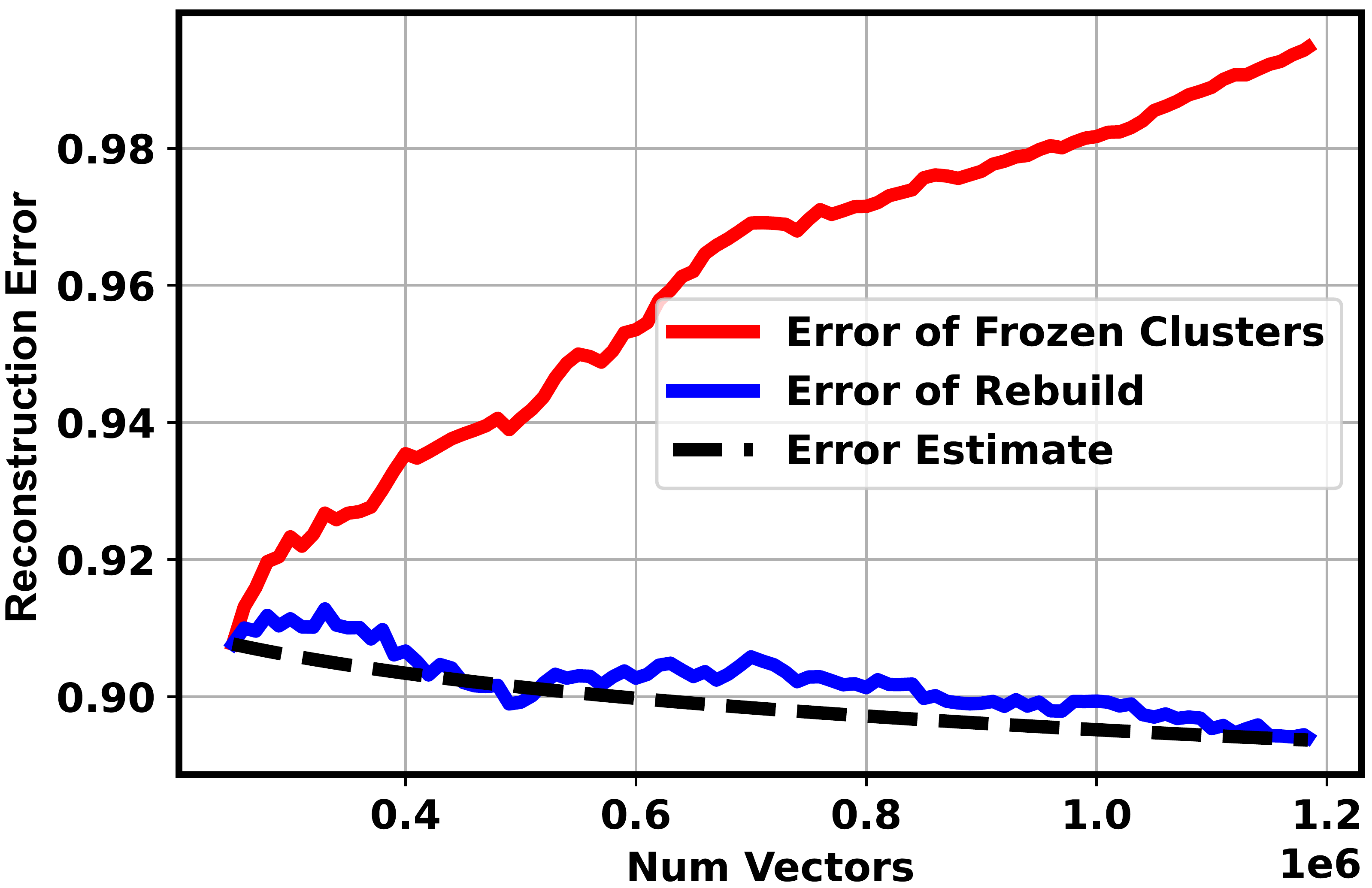}
    \caption{Error estimate $\varepsilon'$ relative to true error for a growing index on Glove1M. Error of frozen clustering grows while the estimate is within $1$\% of that obtained by a full rebuild.}
    \label{fig:glove_error}
\end{figure}

\paragraph{Global Reconstruction Error Function $G_\varepsilon$}

As a part of the global reindexing mechanism, we aim to trigger a rebuild if the global reconstruction error grows high. We propose doing this by estimating the projected improvement in reconstruction error that can be achieved if we perform a full index rebuild. Thus, we define $G_o$ as the relative difference between the current observed error $\varepsilon$ and an ideal error $\varepsilon'$. We define ideal error as the error that would be measured if a full index rebuild were performed, i.e., if k-means were to be applied over the current set of vectors. In other words, if the error of the index deviates far enough from the estimated ideal error, then a rebuild should be performed to recover performance. However, estimating the ideal reconstruction error is challenging and we next describe our approach to address this challenge.

\begin{equation}
G_o = |\varepsilon - \varepsilon'| / \varepsilon'
\end{equation}

\paragraph{Estimating $\varepsilon'$}
An IVF Index can be viewed as a vector quantization scheme, where groups of vectors (clusters) are represented by a representative encoding (centroid). It is proven that k-means clustering produces an empirically optimal quantizer \cite{pollard1981strong} with error $\varepsilon$'. There are well-defined bounds \cite{quantization_bounds} on the difference between the error of an empirically optimal quantizer $\varepsilon$' and an optimal quantizer $\varepsilon^{*}$, where the constant $A$ is a data-dependent parameter.

\begin{equation} \label{eqn:bound_current}
\varepsilon' - \varepsilon^{*} \leq A \sqrt[\leftroot{-2}\uproot{2}d]{\dfrac{1}{n}}
\end{equation}

Similarly, assuming that the initial state of the clustering is empirically optimal, we can derive a bound on its reconstruction error. 

\begin{equation} \label{eqn:bound_init}
\varepsilon_0 - \varepsilon^{*}_0 \leq A \sqrt[\leftroot{-2}\uproot{2}d]{\dfrac{1}{n_0}}
\end{equation}

Combining \ref{eqn:bound_current} and \ref{eqn:bound_init} gives





\begin{equation}
\varepsilon' = \varepsilon_0 \sqrt[\leftroot{-2}\uproot{2}d]{\dfrac{n_0}{n}} + (\varepsilon^{*} - \varepsilon^{*}_0\sqrt[\leftroot{-2}\uproot{2}d]{\dfrac{n_0}{n}})
\end{equation}

We make the assumption that $(\varepsilon^{*} - \varepsilon^{*}_0\sqrt[\leftroot{-2}\uproot{2}d]{\dfrac{n_0}{n}})$ is negligible and achieve the final estimate.

\begin{equation}
\varepsilon' = \varepsilon_0 \sqrt[\leftroot{-2}\uproot{2}d]{\dfrac{n_0}{n}}
\end{equation}


We empirically validate the estimate in Figures \ref{fig:sift1m_error} and \ref{fig:glove_error}. Our experiments show that our estimate is within $2.5$\% and $1$\% of the true error of a rebuild for Sift1M and Glove1M vector datasets, respectively.

%% file: s6_experiments.tex
\section{Experiments}
\label{sec:experiments}

We evaluate Ada-IVF by using a variety of benchmark workloads against state-of-the-art techniques for IVF index maintenance. We begin with an end-to-end performance analysis on an internal workload trace and the BigANN-SS public benchmark (Section \ref{sec:exp-e2e}). We then present an extensive sensitivity analysis to demonstrate the robustness of Ada-IVF compared to baselines across various workload scenarios. To do this, we developed a tool that generates workloads from a given vector dataset with varied read and write properties (Section \ref{sec:sensitivity}). Finally, we present results from a set of microbenchmarks that validate our contributions (Section \ref{sec:microbenchmarks}).

Experimental Highlights
\begin{itemize}[noitemsep,topsep=0pt]
    \item On an internal recommendation workload, Ada-IVF reduces the update time to $62\%$ of that of \emph{LIRE} with a $9\%$ improvement in QPS.
    \item For the public BIGANN-SS benchmark, Ada-IVF reduces the update time by $50\%$ and matches the same QPS as \emph{LIRE}.
    \item Over a comprehensive set of synthetic workload configurations (used for sensitivity analysis), Ada-IVF consistently achieves the highest QPS over IVF baseline methodologies, and compared to \emph{LIRE}, Ada-IVF achieves $1.5-5\times$ higher update throughput across all configurations.
\end{itemize}

\subsection{Experimental Setup}

\begin{table}[t]\footnotesize
    \caption{Workload Characteristics. $r_{rw}=$ read/write ratio, $r_{id}=$ insert/delete ratio}
    \label{tab:workloads}
    \begin{tabular}{c c c c c c }
    \toprule
    Workload & Update Size & $r_{rw}$ & $r_{id}$ & Write Locality & Read Locality\\
    \midrule
    Internal & $\approx$ 2\% &  10.0 & $\approx5.0$  & high & high \\
    BigANN-SS & 0.001\% - 10\% & .01 & $\approx2.0$ & high & high \\
    Generator & variable & variable & variable & variable & variable \\
    \bottomrule
    \end{tabular}
\end{table}

\subsubsection{Workloads}
\label{sec:setup-real-world}

\textbf{Internal} is a real-world, industrial workload for an online recommendation application where both the updates to the index and the queries are processed in a streaming fashion. This read-heavy workload exhibits significant write locality as each batch of updates primarily targets a specific partition. We execute this workload using access properties derived from real-world data using MSTuring10m as the underlying vector dataset.

We also use \textbf{BigANN-SS}, a \emph{batch} public streaming benchmark that is developed as part of the Neurips'23 BigANN competition. BigANN-SS uses a 30M subset of the MSTuring dataset, and its workload consists of batch inserts and delete,s where each batch is based on a cluster of the original dataset. By inserting and deleting based on partitions, BigANN-SS's workload exhibits update locality. The inserts and deletes vary in size between 10-250k. After each update, the same set of 10k queries are evaluated in bulk with a target recall of 0.9. 

\subsubsection{Workload Generator}
\label{sec:workload_simulator}

We use a configurable workload generator that can simulate a variety of settings on any vector dataset.
Using such a generator helps us evaluate the sensitivity and robustness of Ada-IVF and its components on public vector datasets.
Given a vector dataset and workload parameters, the generator clusters the dataset and samples from the clusters to produce inserts, deletes, and queries. The six primary parameters of the generator are (i) the initial size $s_0$, (ii) the update size $s_u$, (iii) the insert/delete ratio $r_{id}$, (iv) the update cluster sample fraction $CSF_u$, (v) the read/write ratio  $r_{rw}$ and (vi) the query cluster sample fraction $CSF_q$, which is optional if queries are provided with the dataset. The update and query sample fractions allow for configuring the \emph{locality} of updates and queries by controlling the fraction of vectors sampled from a given cluster to produce the update/query. With CSF=1.0, an entire cluster of vectors is sampled for an insert/delete; therefore, the update is highly localized. A small CSF (e.g., CSF=.001) corresponds to a less localized update, as vectors are sampled from many clusters across the vector space. We use the workload generator to conduct a sensitivity analysis using the MSTuring10M dataset. We use Sift1M and Glove1M for microbenchmarks. Unless otherwise specified, for all cases, we fix the following parameters:  $s_0 = 0.1|X|$, $s_u = 10000$, $r_{id}=\inf$ (insert only), $CSF_u = 1.0$, $r_{rw}=0.1$, and use queries that come with the dataset so $CSF_q$ does not apply. 

\begin{figure}[]
    \centering
    \includegraphics[width=.5\textwidth]{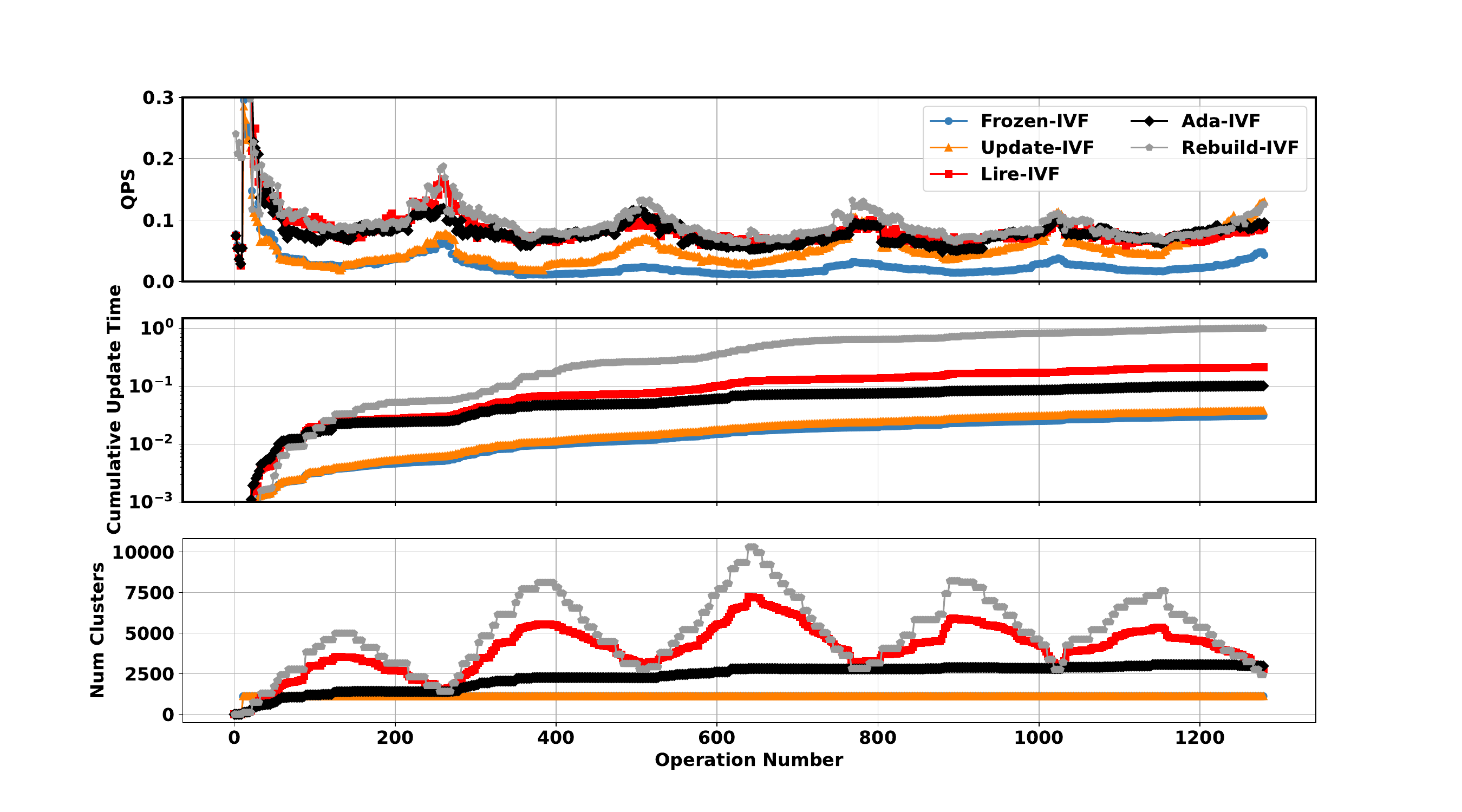}
    \caption{Workload trace for the BigANN-SS benchmark. Recall Target = .9. QPS and update times are normalized to the maximum observed value over the trace.}
    \label{fig:bigannss_trace}
\end{figure}

\begin{table}[t]\small
    \caption{Normalized search and update throughput for each workload. Throughput is normalized by the values obtained by \emph{Rebuild}. Recall Target = .9.}
    \label{tab:e2e_workloads}
    \begin{tabular}{c c C{2.5cm} C{2.5cm}}
        \toprule
        Workload & Index & Normalized Search Throughput & Normalized Update Throughput \\
        \cmidrule{1-4}
        \multirow{5}{*}{Internal} 
        & Frozen & .81 & 20.0 \\
        & Update & .84 & 12.5 \\
        & LIRE & .83 & 4.2 \\
        & \textbf{Ada-IVF} & $\mathbf{.89}$ & $\mathbf{6.7}$ \\
        & Rebuild & 1.0 & 1.0 \\
        \cmidrule{1-4}
        \multirow{5}{*}{BigANN-SS} 
        & Frozen & .27 & 33.3 \\
        & Update & .54 & 25.0 \\
        & LIRE & .89 & 4.7 \\
        & \textbf{Ada-IVF} & $\mathbf{.85}$ & $\mathbf{10.0}$ \\
        & Rebuild & 1.0 & 1.0 \\
        \bottomrule
    \end{tabular}
\end{table}

\begin{figure*}[!tp]
    \centering
    \begin{subfigure}[b]{0.24\textwidth}
        \includegraphics[width=\linewidth]{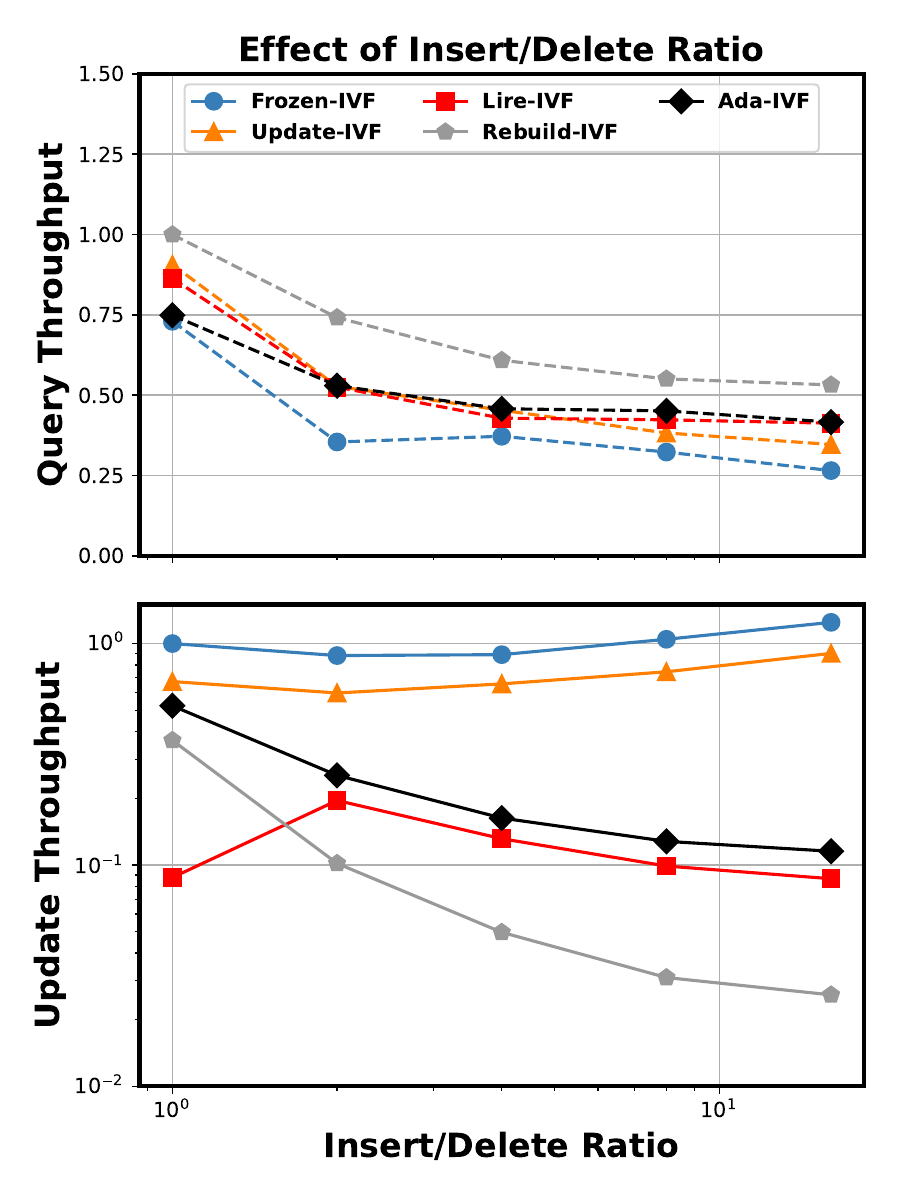}
        \caption{Vary Insert/Delete Ratio}
        \label{fig:id_ratio}
    \end{subfigure}
    \hfill 
    \begin{subfigure}[b]{0.24\textwidth}
        \includegraphics[width=\linewidth]{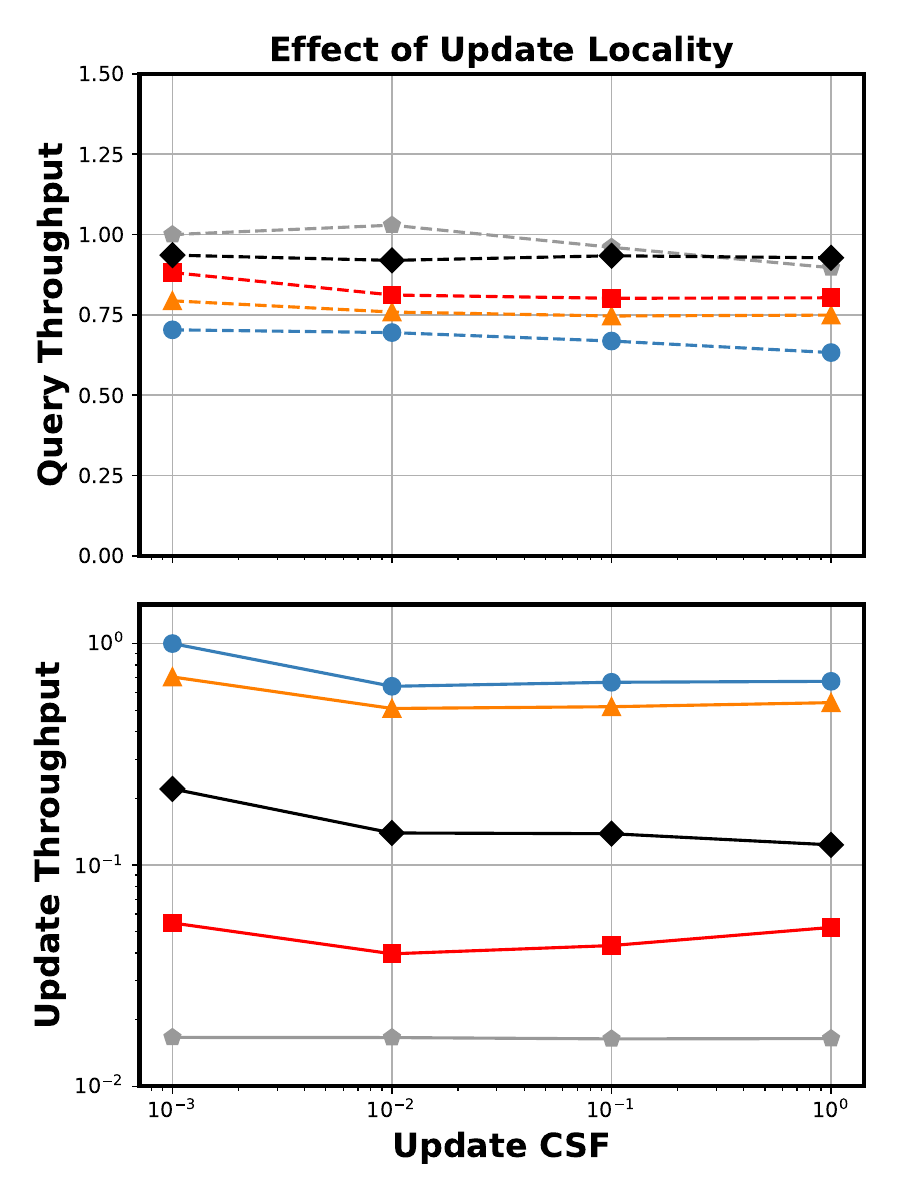}
        \caption{Vary Update Locality}
        \label{fig:update_locality}
    \end{subfigure}
    \hfill 
    \begin{subfigure}[b]{0.24\textwidth}
        \includegraphics[width=\linewidth]{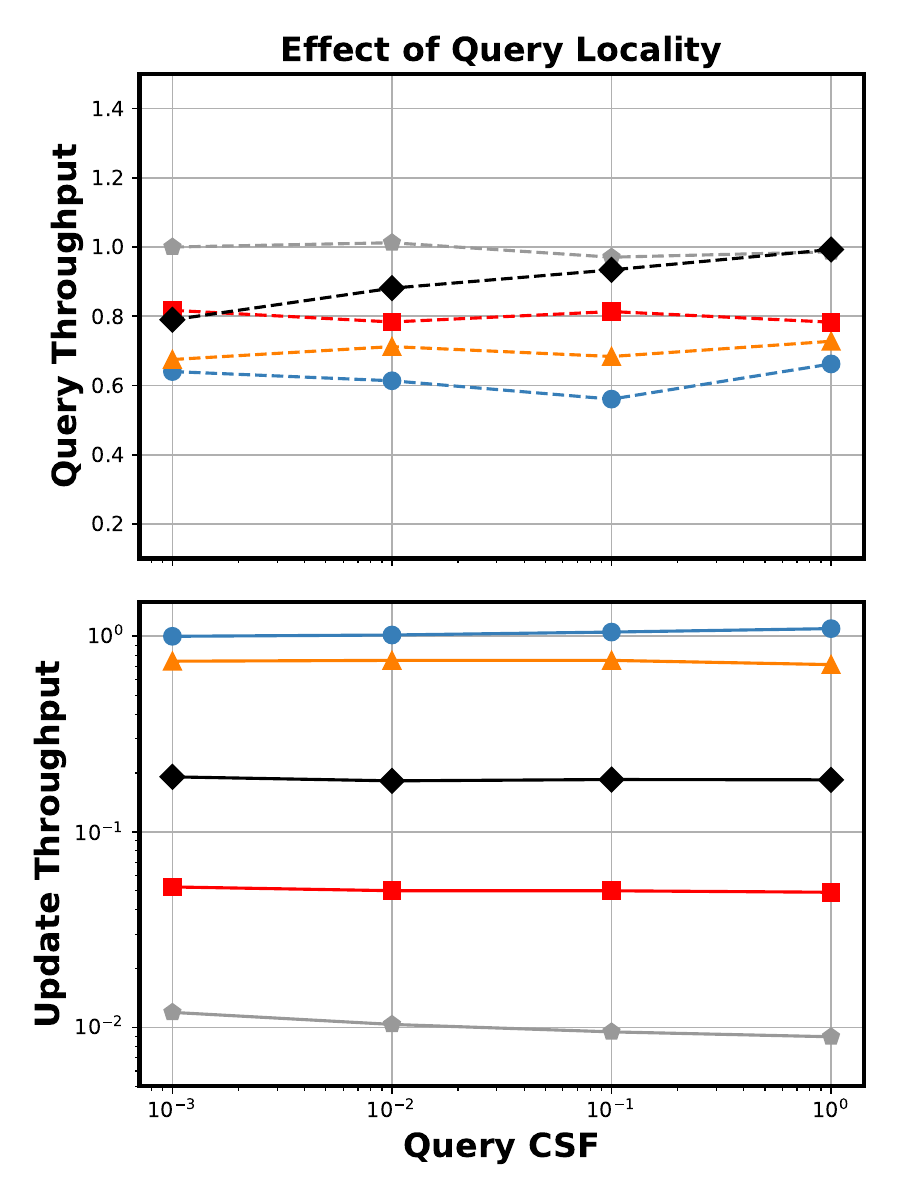}
        \caption{Vary Query Locality}
        \label{fig:query_locality}
    \end{subfigure}
    \hfill 
    \begin{subfigure}[b]{0.24\textwidth}
        \includegraphics[width=\linewidth]{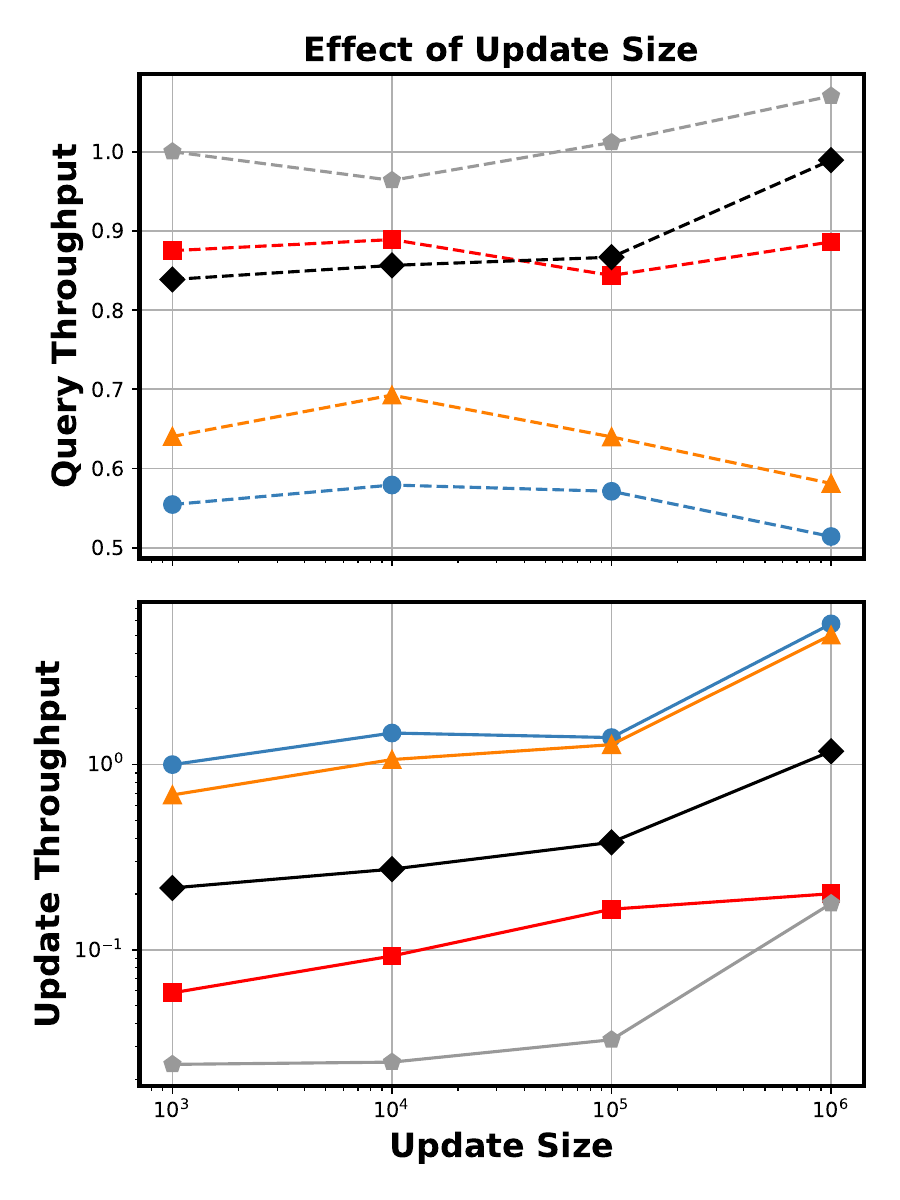}
        \caption{Vary Update Size}
        \label{fig:update_size}
    \end{subfigure}
    \caption{Sensitivity analysis conducted using the workload generator. Each experiment varies a single parameter and measures the effect on search and update throughput. Throughput is normalized to the maximum observed value for each experiment.}
    \label{fig:unified_label}
\end{figure*}

\subsubsection{Baselines}

We compare Ada-IVF's performance against the following standard and state-of-the-art maintenance strategies that are employed in existing systems for handling updates in IVF Indexes (Section \ref{sec:background}). Ada-IVF's configurable indicator functions and flexible reindexing manager allow us to implement these algorithmic baselines in our system easily. This ensures that the system overheads are the same for all strategies. For ease of explanation, queries are serially executed in a synchronous fashion for all approaches and workloads.
\begin{itemize}
    \item \emph{Frozen}: Does no maintenance.
    \item \emph{Update} \cite{baranchuk_dedrift_2023, arandjelovic2013all}: Updates centroids on partition modification to reflect the true mean of the partition.
    \item \emph{Rebuild} \cite{wei2020analyticdb}: Periodically rebuilds the index.
    \item \emph{LIRE} \cite{xu_spfresh_2023}: Performs splitting and merging of partitions if a target size is violated. After splitting and merging, vector reassignment of neighboring partitions is performed. 
    LIRE is implemented within Ada-IVF with the corresponding parameters $\tau_f = 0$, $\iota = 0$, and $\beta=1.0$.
\end{itemize}

We set the target partition size for all baselines and workloads to 1000, which obtains good performance across workloads. We match the shared configuration parameters of Ada-IVF and \emph{LIRE}, setting the maximum/minimum partition size to 2000/500 and using a reindexing radius of 25. For \emph{Rebuild}, we trigger rebuilding after $2.5\%$ of the total number of vectors in the workload have been modified. We individually tune the Ada-IVF specific parameters $\tau_f$, $\alpha$, and $\beta$ for each workload. We measure normalized search throughput at a target recall of .9 and normalized update throughput. For clarity, we refer to search throughput as QPS. We run all experiments using an R5-24xlarge AWS instance with 96 vCPUs and 768GB of memory.

\subsubsection{Implementation} We implement Ada-IVF using C++/Python using PyTorch 2.0 \cite{DBLP:journals/corr/abs-1912-01703} as the underlying tensor library. OpenMP 5.1 \cite{660313} is used for parallelizing index operations. We use PAM \cite{Sun_2018} to map vector IDs to partition IDs
for efficient deletes.
\subsection{Industrial and Benchmark Workloads}
\label{sec:exp-e2e}


Table \ref{tab:e2e_workloads} shows the search and update throughput on Internal and BigANN-SS of Ada-IVF and baselines relative to \emph{Rebuild}, which is optimal for search throughput. Below, we detail each workload.

\paragraph{BigANN-SS}

Figure \ref{fig:bigannss_trace} shows the workload trace for the BigANN-SS benchmark. In this workload, a series of inserts and deletes are performed, which grow and shrink the set of vectors in the index over time. First, looking at the search QPS (top figure), we observe that \emph{Frozen} and \emph{Update} achieve 3.5$\times$ and $1.9\times$ less QPS than \emph{Rebuild}, while \emph{LIRE} and Ada-IVF can achieve similar QPS. We attribute this to the re-balancing mechanisms applied by both methods. There is still a gap between the QPS of Ada-IVF and \emph{Rebuild}, suggesting room for improvement. The middle figure shows the cumulative update time over the workload normalized by \emph{Rebuild}'s update time. \emph{Frozen} and \emph{Update} achieve the lowest update times as they have minimal overhead. \emph{LIRE} and Ada-IVF obtain an update throughput of $4.7\times$ and $10.0\times$ relative to \emph{Rebuild}, resulting in a lower total update time. Ada-IVF achieves a high update throughput due to the indicator function it employs, as it is not as aggressive in selecting partitions for reindexing. This can be observed in the bottom figure, which shows the number of partitions over time. The number of partitions varies significantly for \emph{LIRE}, while Ada-IVF remains more stable throughout the workload. Overall, Ada-IVF achieves a $2\times$ higher update throughput over \emph{LIRE} while matching query throughput. Compared to \emph{Rebuild}, it achieves $10\times$ higher update throughput with $85\%$ of the search throughput.

\paragraph{Internal Workload}

Table \ref{tab:e2e_workloads} shows the search and update throughput for Internal. We see that \emph{Frozen} and \emph{Update} perform relatively well, showing an order of magnitude higher update time compared to \emph{Rebuild} with only a $.81\times$ and $.84\times$ reduction in QPS for \emph{Frozen} and \emph{Update}, respectively. We attribute the good performance of the \emph{Frozen} and \emph{Update} to the large initial size of the workload. Unlike BigANN-SS, Internal starts with a significantly larger set of vectors, and \emph{Frozen} and \emph{Update} obtain a more representative initial clustering of the data points. \emph{LIRE} achieves a similar QPS as \emph{Frozen} with $4.7\times$ higher update throughput than \emph{Rebuild}. Ada-IVF achieves a higher QPS than baseline methods with $.89\times$ QPS with a further $1.6\times$ improvement in update throughput over \emph{LIRE}. 

\subsection{Sensitivity Analysis}
\label{sec:sensitivity}


\begin{figure*}[!htb]
\minipage{0.32\textwidth}
    \centering
    \includegraphics[width=\textwidth]{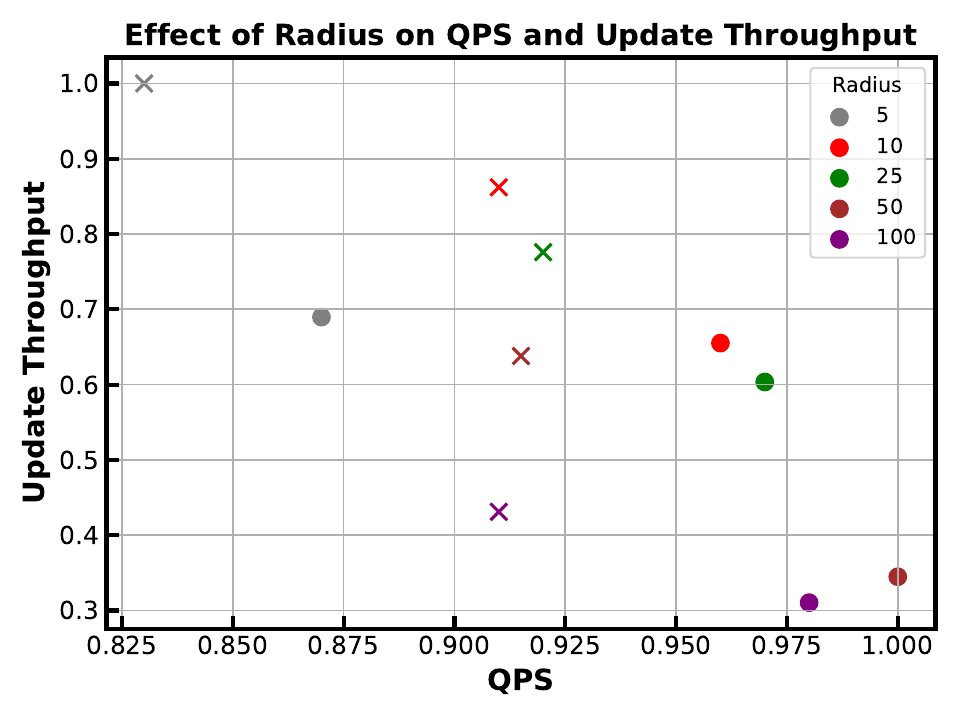}
    \caption{Parameter sweep varying the reindexing radius $=$ \{5, 10, 25, 50, 100\}. X denotes \emph{LIRE} configurations, and search and update throughput are normalized to the maximum value.}
    \label{fig:reindexing_radius}
\endminipage\hfill
\minipage{0.32\textwidth}
    \centering
    \includegraphics[width=\textwidth]{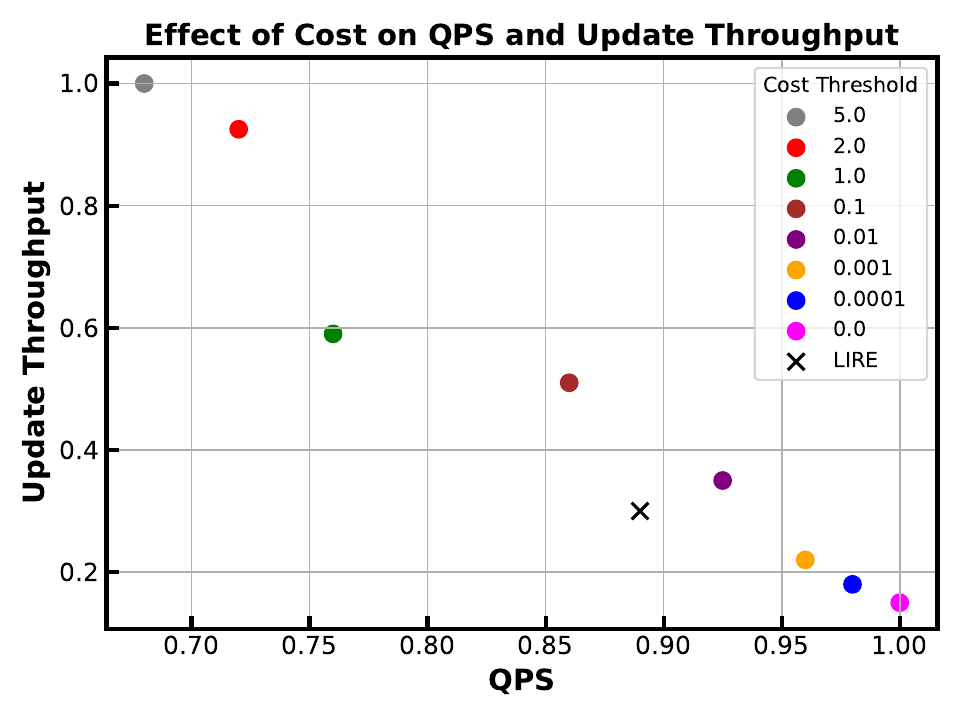}
    \caption{Parameter sweep varying the indicator threshold $\tau_f$ $=$ \{0, .0001 .001, .01, .1, 1.0, 2.0, 5.0\}. X denotes \emph{LIRE} configurations, and search and update throughput are normalized to the maximum value.}
    \label{fig:cost_threshold}
\endminipage\hfill
\minipage{0.32\textwidth}%
    \centering
    \includegraphics[width=\textwidth]{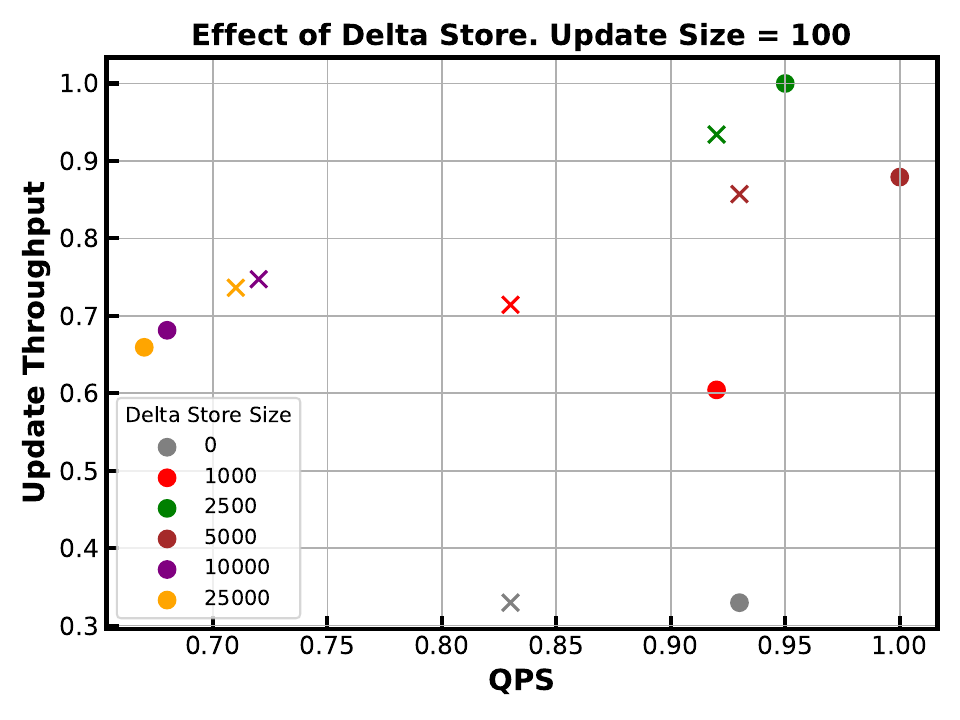}
    \caption{Parameter sweep varying the delta store size $=$ \{0, 1000, 2500, 5000, 10000, 25000\}, X denotes \emph{LIRE} configurations, and search and update throughput are normalized to the maximum value.}
    \label{fig:delta_store}
\endminipage
\end{figure*}

To demonstrate the robustness of Ada-IVF to a variety of workload scenarios, we conduct a sensitivity analysis of Ada-IVF and baselines update methodologies using a workload generator. For each experiment, we vary a single generator parameter while keeping the rest fixed, as discussed in \ref{sec:workload_simulator}. 

\paragraph{Insert/Delete Ratio}

First, we vary the insert/delete ratio, which determines the index's rate of growth. We see that Ada-IVF matches the query performance of \emph{LIRE} with a slight improvement in update time as the ratio increases. We also observe a general decrease in update and query throughput as the ratio increases. This is due to the growing dataset size for positive ratios.

\paragraph{Update Locality}

Here, we vary the cluster sample fraction (CSF) used to generate inserts and deletes. This controls the locality of updates (\ref{sec:workload_simulator}). Figure \ref{fig:update_locality} shows the query and update throughput for Ada-IVF and baseline methods. Ada-IVF consistently matches the QPS of a \emph{Rebuild} across all sample fractions, while other methods show degradation in QPS. \emph{Frozen} and \emph{Update} achieve the highest update throughput, with Ada-IVF achieving the next best update throughput, a $2-5\times$ improvement over \emph{LIRE}. We see that results are consistent across sample fractions.

\paragraph{Query Locality}

Now, we vary the cluster sample fraction (CSF) used to generate queries. As query locality increases, Ada-IVF increases its QPS while maintaining a consistent update throughput ($4\times$ \emph{LIRE}). This suggests that the temperature component causes aggressive reindexing of only the often-scanned partitions and thus increases query throughput. Other methods exhibit consistent behavior across all cases, which is expected given they do not consider locality. Overall, this validates our temperature model for utilizing read locality information.

\paragraph{Update Size} Figure \ref{fig:update_size} shows the effect of varying the update size on index performance. First, observing the impact on read throughput, we see Ada-IVF and \emph{LIRE} obtain similar query performance across update sizes while \emph{Frozen} and \emph{Update} degrade in performance as the update size increases. For update throughput, all methods show an increase in throughput with increasing update size, demonstrating the effectiveness of batch on IVF index updates. 

\paragraph{Takeaways}

We see that across most workload scenarios, Ada-IVF best matches the query performance of \emph{Rebuild} at a fraction of the update time of the next best baseline \emph{LIRE}. Because of read temperature, Ada-IVF is especially powerful for workloads that exhibit significant read locality, where query performance increases as locality increases.

\subsection{Microbenchmarks}
\label{sec:microbenchmarks}

We now evaluate individual components of our design in further detail using a synthetic SIFT1M workload with $r_{id} = 1$. We first evaluate the effect of indicator function parameters by performing a comprehensive parameter sweep of Ada-IVF parameters. We then perform a similar parameter scan to evaluate the effect of varying the local reindexing mechanism parameters. We then evaluate the effect of batching updates.

\paragraph{Effect of Reindexing Radius $r_c$}

Figure \ref{fig:reindexing_radius} shows the effect of varying the reindexing radius. \emph{LIRE} configurations ($\tau_f$ $=0$, $\iota=0$, and $\beta=1.0$) are marked with an X. We see that tuning the reindexing radius provides a clear trade-off between update throughput and query performance. We also observe that \emph{LIRE} has higher update throughput for the same radius but with a lower QPS than Ada-IVF. \emph{LIRE} also shows minimal QPS improvement beyond a radius of 10. Generally, a larger reindexing radius for Ada-IVF increases QPS and decreases update throughput. Index maintainers can tune this parameter to best meet their workload's needs.

\paragraph{Effect of Local Indicator Threshold $\tau_f$}

Figure \ref{fig:cost_threshold} shows the effect of varying the threshold $\tau_f$, which determines the value of $f$ needed for a violator to be reindexed. \emph{LIRE} configurations are marked with an X. Like the reindexing radius experiment, we see that tuning the local indicator threshold provides a clear trade-off between update and query throughput. A smaller threshold will trigger reindexing more often and better maintain index performance at a cost in update throughput. Generally, a smaller threshold increases QPS and decreases update throughput. 



\paragraph{Delta Store}

Now, we evaluate the effect on index performance when batching inserts with a \emph{delta store}. The delta store is a fixed capacity list that stores recent updates. The contents of the delta store are scanned exhaustively by searches. When the delta store reaches capacity, its contents are flushed to the IVF index.
Figure \ref{fig:delta_store} shows the effect of varying the capacity of the delta store for a small update scenario (update size = 100).  As before, we mark \emph{LIRE} configurations with an X. We see that including a delta store can significantly improve update throughput and observe a slight improvement in QPS for small update sizes. However, if the delta store grows too large, query performance significantly degrades as queries must scan the delta store exhaustively. In conclusion, maintainers operating in small update environments can benefit significantly from the inclusion of a delta store in their index.


%% file: s7_discussion.tex
\section{Discussion}
\label{sec:discussion}
We outline extensions of our method to Product Quantization and hierarchical IVF indexes, and then we discuss related work.

\subsection{Extensions}

The IVF index maintenance techniques introduced in this work apply to single-level IVF indexes, but we believe they are applicable to any type of index based on vector quantization. Below, we discuss extensions of our index maintenance technique to two commonly used indexes: Product Quantization \cite{jegou_product_2011} and Hierarchical IVF indexes \cite{sun_automating_2023}.

Product Quantization (PQ) encodes a d-dimensional vector into a d'-dimensional code vector using $d'$ codebooks. Codebooks are obtained by KMeans to cluster $d'$ subspaces of the vector distribution. Product quantization is generally used for lossy vector compression \cite{jegou_product_2011} and is a commonly used technique in ANN index design \cite{wei2020analyticdb}. In a dynamic setting where new vectors are added, and existing vectors are deleted, the quality of the codebook trained on the initial vector dataset decreases. This eventually necessitates an expensive codebook rebuild and re-encoding of vectors to recover index quality.
In contrast with IVF indexes, imbalance does not negatively affect query performance, as the only objective is to minimize the quantization error of the encoding. Hence, when applied in this setting, our indicator functions (Section \ref{sec:iqm}) should only consider the error term ($\beta = 0$, $\gamma = 0$). However, a significant challenge with applying our method to PQ quantization is that, typically, there are few clusters in a given codebook. For example, it's common to use 8-bit codes corresponding to a total of 255 centroids in each codebook \cite{jegou_product_2011}, meaning that a modest reindexing radius of 25 will recluster $10\%$ of the data if a single cluster violates the indicator function. Therefore, local reindexing is more applicable to cases where larger codes (e.g., 16-bit codes, 65k centroids) are used. There exists some work that studies the impact of updates in the context of product quantization \cite{xu2018online, liu2020online}. Future study is required to assess the performance implications of using Ada-IVF's maintenance technique over product quantization codebooks.

Hierarchical IVF indexes \cite{sun_automating_2023} efficiently scale to large data sizes by maintaining a hierarchical KMeans clustering. In a basic two-level structure, vectors are grouped into fine-grained clusters whose centroids are further aggregated into coarse clusters. Changes to underlying vectors necessitate reindexing these fine clusters, impacting the coarse-level structure by centroid updates. We can extend our single-level methodology in Ada-IVF by using an independent update manager for each hierarchy level. This ensures that any changes in the fine-grained centroids have a limited effect on the quality of the coarse-grained clustering. Utilizing this approach, we can extend our method to maintain hierarchical IVF indexes.

We reserve the study of streaming workloads and approaches for handling updates on other types of indexes for future work.

\subsection{Related Work}

\paragraph{Vector Indexes}
Indexes for vector search predominately fall under two categories: partitioned and graph indexes. IVF indexes are the most prominent type of partitioned index with a multitude of variants \cite{wei2020analyticdb, ferrari_revisiting_2018, jegou2011searching, guo_accelerating_2020, chen_spann_nodate, babenko_inverted_2015}. IVFADC \cite{jegou2011searching} pairs IVF indexes with Product Quantization (PQ). SCANN uses a hierarchical IVF index and PQ and includes optimizations for Maximum Inner Product Search (MIPS). \cite{guo_accelerating_2020, sun_automating_2023}. SPANN \cite{chen_spann_nodate} replicates vectors across partitions and includes a graph index over the centroids to accelerate centroid scans. Hash-based \cite{jain_online_2008, sundaram_streaming_2013, gionis_similarity_nodate}, learned \cite{li_learning_2023, mamun_survey_2024, gupta_bliss_2022, gupta_elias_nodate}, and tree-based \cite{CHEN2019145, wang_graph_nodate, li_constructing_2023, lin_ann-tree_2001, omohundro_five_nodate} indexes are alternative types of partitioned indexes, but do not yet outperform leading IVF and graph approaches on public benchmarks \cite{aumüller2018annbenchmarks, simhadri_results_2022}. Graph indexes \cite{ni_diskann_2023, wang_starling_2024, xu2022proximity, fu_fast_2018} are the leading alternative solution to IVF indexes; however, they face challenges supporting updates at scale due to their large memory overhead and random access patterns \cite{xu_spfresh_2023}. SPFresh (the system that implements LIRE) shows superior performance to the leading updatable graph index DiskANN \cite{singh2021freshdiskann, ni_diskann_2023}. 

\paragraph{Vector Data Management Systems}
There has been an explosion of interest in vector search due to the success of embedding-based machine learning methods. To meet demand, new vector data management systems, known as vector databases, have sprouted \cite{wang2021milvus, noauthor_vald_nodate, noauthor_qdrant_nodate, noauthor_vespa_nodate, noauthor_ai-native_nodate, noauthor_why_nodate, noauthor_welcome_nodate, noauthor_vector_nodate} and existing data management systems have added vector search capability \cite{noauthor_pgvectorpgvector_nodate}. Vector databases are typically general to the underlying index and support both IVF and graph index approaches. We believe our maintenance approach can be directly adopted by vector databases to improve their performance in update-heavy environments.

\paragraph{Updatable Spatial Indexes}
The R-tree \cite{guttman_r-trees_1984} is a classic example of a dynamic spatial index. R-trees maintain index performance by splitting and deleting leaf nodes once they become full or too small. The mechanism for splitting nodes aims to minimize overlap between the resulting nodes such that queries visit as few nodes as possible. In IVF indexes, overlap between clusters is difficult to measure, so we instead use the reconstruction error. Numerous r-tree variants have been proposed to improve the splitting procedure by minimizing overlap, such as \cite{beckmann_r-tree_1990} and \cite{berchtold_x-tree_1996}.

%% file: s8_conclusion.tex
\section{Conclusion}
We introduced Ada-IVF, an incremental methodology for maintaining IVF index performance for dynamic workloads. Ada-IVF uses local and global indicator functions that determine which clusters need to be reindexed using local reindexing. We evaluated our approach across a variety of benchmarks and found that compared with the state-of-the-art dynamic IVF index maintenance strategy, Ada-IVF achieves an average of $2\times$ and up to $5\times$ higher update throughput while matching query throughput across a range of benchmark workloads.